\documentclass[aps,prb,twocolumn,floatfix,amsmath,amssymb]{revtex4}
\usepackage[final]{graphicx}
\usepackage{bm}
\usepackage{afterpage}
\usepackage{color}
\usepackage{comment}
\usepackage[colorlinks=true,urlcolor=blue,linkcolor=blue,citecolor=blue]{hyperref}

\emergencystretch=\maxdimen
\begin{document}

\title{Magnon Landau levels in the strained antiferromagnetic honeycomb nanoribbons}

\author{Junsong Sun}
\affiliation{School of Physics, Beihang University,
Beijing, 100191, China}

\author{Huaiming Guo}
\email{hmguo@buaa.edu.cn}
\affiliation{ School of Physics, Beihang University,
Beijing, 100191, China}
\affiliation{Beijing Computational Science Research Center, Beijing 100193, China}

\author{Shiping Feng}
\affiliation{ Department of Physics,  Beijing Normal University, Beijing, 100875, China}

\begin{abstract}
The pseudo-magnetic field created by a non-uniform unaxial strain is introduced into the antiferromagnetic honeycomb nanoribbons. The formation of magnon pseudo-Landau levels, which appear from the upper end of the spectrum and whose level spacings are proportional to the square root of the level index, is revealed by the linear spin-wave theory. The antiferromagnetic order is gradually weakened along the $y$-direction by the strain. At large enough strength, the system is decoupled into isolated zigzag chains near the upper boundary, and demonstrates one-dimensional magnetic property there. While the quantum Monte Carlo simulations also predict such a transition, this exact method gives a critical point deeper in the bulk. We also investigate the $XY$ antiferromagnetic honeycomb nanoribbons, and find similar pseudo-Landau levels and antiferromagnetic evolution. Our results unveil the effect of a non-uniform unaxial strain on the spin excitaions, and may be realized experimentally based on two-dimensional quantum magnetic materials.
\end{abstract}


\maketitle

%% \textit{Introduction.-}
\section{Introduction}

The mechanical strain has become a powerful tool to engineer the electronic property of graphene and other two-dimensional quantum materials~\cite{PhysRevLett.103.046801,Guinea2010Energy,2010Strain,AMORIM20161}. The low-energy physics of graphene is described by the Dirac Hamiltonian, in which the perturbation of a strain acts as a vector potential with opposite signs at the two valleys~\cite{PhysRevB.87.165131,RevModPhys.81.109}. Experimentally, a controlled uniaxial strain can be readily realized in graphene using feasible techniques. However such a strain results in a constant gauge field, which shift the position of the Dirac points in opposite directions and can only induce a band gap at unrealistic large strength~\cite{PhysRevB.80.045401,PhysRevB.79.205433}.

A nonzero pseudo-magnetic field (PMF), especially a uniform one which can mimic the effect of a real magnetic field, is highly desirable. Since the PMF magnitude is proportional to the gradient of the strain, a PMF should be created by a non-uniform strain~\cite{PhysRevB.88.115428,PhysRevB.93.035456}. Guinea et al. first predict a triaxial strain can lead to a strong uniform PMF~\cite{Guinea2010Energy}. Later on, experimentally more available approaches, such as bending or twisting graphene~\cite{PhysRevB.81.035408,PhysRevB.86.125402,PhysRevLett.112.096805,PhysRevB.96.155446}, are proposed to generate an almost uniform PMF. While it is still challenging to directly realize the above theoretical proposals, the pseudo-Landau levels (PLLs) induced by PMFs have been observed by scanning tunneling microscopy in highly localized regions of graphene with non-planar deformations~\cite{2010natureStrain,RN45,PhysRevB.87.205405}.

Recently, strain-induced gauge fields have been generalized to three-dimensional Dirac and Weyl semimetals~\cite{PhysRevX.6.041021,PhysRevB.96.224518,PhysRevB.95.041201}, and even neutral quasiparticles such as: Bogoliubov particles in two-dimensional nodal superconductors~\cite{PhysRevB.97.024520}, magnons in honeycomb antiferromagnets~\cite{PhysRevLett.125.257201,PhysRevB.103.144420}, et al.~\cite{PhysRevLett.116.167201,PhysRevA.103.013505,PhysRevLett.118.194301}. Although Landau quantization of neutral quasiparticles is formed by the strain-induced PMF, its properties are not completely identical to that in graphene. Specifically, under PMF induced by the triaxial strain in a honeycomb antiferromagnet, the PLLs appear at the upper end of the magnon spectrum, and are equally spaced~\cite{PhysRevLett.123.207204,sun2021quantum}. The strain is introduced to the honeycomb antiferromagnet by analogizing the exchange coupling as the hopping amplitude in graphene and modifying it in the same way as the latter. The strain-induced gauge field may not act on the magnons exactly as it does in graphene. Hence it is natural to ask whether the different methods used to engineer a uniform PMF in graphene have the same effect on the spin excitations in honeycomb antiferromagnets.

\begin{figure}[htbp]
\centering \includegraphics[width=9.cm]{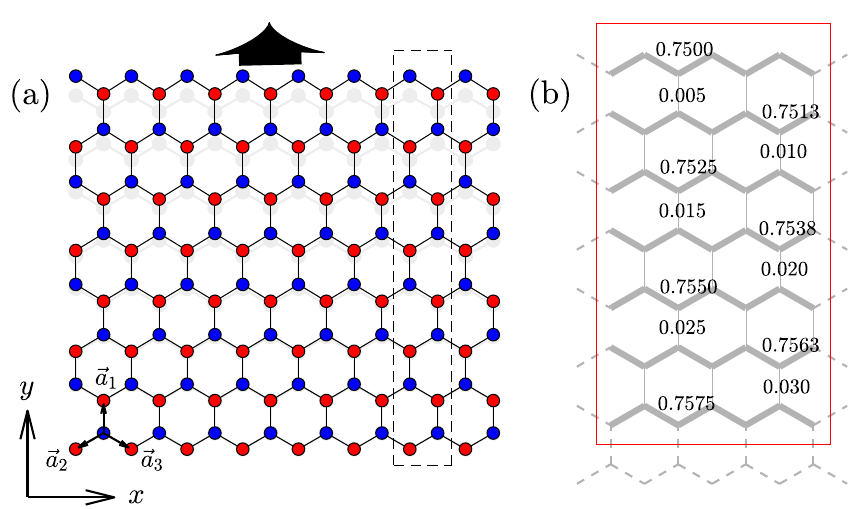} \caption{(a) Schematic representation of a strained honeycomb nanoribbon with zigzag boundaries.
The nanoribbon is periodic along the $x$-direction, and the width in the $y$-direction is $L_y=8$ for the figure. (b) Enlarged plot in the vicinity of the upper boundaries of a $L_y=200$ honeycomb nanaribbon. The value of the exchange coupling on each bond is represented by the thickness of the bond and marked explicitly near the bonds. $c/c_{max}=1$ is used for the strain strength in (b).}
\label{fig1}
\end{figure}

In this manuscript, we study the Heisenberg and $XY$ Hamiltonians on the honeycomb lattice under a non-uniform unaxial strain utilizing the linear spin-wave theory (LSWT) and quantum Monte Carlo (QMC) method. First, the formation of PLLs is revealed by LSWT. Then we address the evolution of the antiferromagnetic (AF) order, which is characterized by a finite local magnetization and long-range AF correlations. Both approaches show the AF order is reduced monotonically in the $y$-direction, and predict a critical position, beyond which the system is described by decoupled one-dimensional Heisenberg chains, for large enough strain strength. Finally, we present the results of the $XY$ Hamiltonian under the same kind of strain. Here the formation of PLLs and the evolution of the AF order with the strain is very similar to the Heisenberg case, except that the magnetic order is more robust and persists even at the largest possible strain strength. These results are closely related to the two-dimensional quantum magnetic materials, and will attract both theoretical and experimental interests.

This paper is organized as follows. Section II introduces the precise
model we will investigate, along with our computational methodology.
Section III presents the magnon Landau levels in LSWT. Section IV uses LSWT and QMC simulations to study the evolution of the AF order. Section V demonstrates the results of the $XY$ Hamiltonian under a unaxial strain. Section VI contains some further discussions and the conclusions.

\section{The model and method}

We consider an AF Heisenberg model on a honeycomb nanoribbon, which in the absence of strain writes as
\begin{align}
H_0=J\sum_{\langle ij\rangle}{\bf S}_i\cdot{\bf S}_j,
\end{align}
where $J$ is the AF exchange coupling; ${\bf S}_{i}=(S_i^{x},S_i^{y},S_i^{z})$ is spin-$\frac{1}{2}$ operator on the site $i$, which obeys commutation relations, $\left[S^{\mu}_{i},S^{\nu}_{j}\right]=i\hbar \varepsilon _{\mu\nu\tau}S^{
\tau}_{i}\delta_{ij}$ with $\varepsilon _{\mu\nu\tau}$ the Levi-Civita symbol and $\mu,\nu,\tau= x,y,z$ representing spin components.
In the presence of strain, the lattice is deformed and the Hamiltonian is modified through a simple modulation of the exchange couplings. For small displacements, we have
\begin{align}
J\longrightarrow J_{ij}=J(1-\gamma \Delta u_{n}),
\end{align}
where $\gamma$ represents the strength of magnetoelastic coupling; $\Delta u_{n} (n=1,2,3)$ is the relative displacement of the bond, given by
\begin{align}
\Delta u_{n}=\sum_{i, j} \frac{a_{n}^{i} a_{n}^{j}}{a_{0}^{2}} \epsilon_{i j}.
\end{align}
In the above equation, $\vec{a}_{n}$ are the nearest-neighbor vectors, and the strain tensor is $\epsilon_{i j}=\frac{1}{2}\left[\partial_{j} u_{i}+\partial_{i} u_{j}\right]$ $(i,j=x,y)$ with the displacement $\vec{u}({\bf r})=[u_x({\bf r}),u_y({\bf r})]$ of the lattice site at position ${\bf r}=(x,y)$. Here $\vec{u}({\bf r})$ is assumed to depend only on $y$, which results $\epsilon_{xx}=\epsilon_{xy}=\epsilon_{yx}=0$ and $\Delta u_1=\epsilon_{yy}, \Delta u_2=\Delta u_3=\epsilon_{yy}/4$~\cite{PhysRevB.101.085423}.
The strain tensor is expected to generate a pseudo-gauge field
\begin{align}
\mathbf{A}=\frac{\gamma}{2}\left(\begin{array}{c}
\epsilon_{x x}-\epsilon_{y y} \\
-2 \epsilon_{x y}
\end{array}\right).
\end{align}
For our case, only $A_x=-\frac{\gamma}{2}\epsilon_{y y}$ is nonzero.
We take $\epsilon_{y y}=\frac{c}{\gamma}y$, which generates a homogeneous pseudo-magnetic field $\vec{B}=\frac{1}{2}c\hat{z}$. We set the bottom of the ribbon as the coordinate origin of the $y$-direction, thus $y_j=\frac{3}{2}(j-1)+\frac{1}{2}$ for the blue atoms in the $j$-th zigzag horizontal chain (see Fig.\ref{fig1}). Since the exchange coupling of the vertical bond decreases more rapidly with $y$, the maximum strain parameter $c_{max}$ is determined by the appearance of zero exchange coupling on the bonds along the $y$-direction. For a ribbon with fixed width $L_y$, the vertical bonds connecting the blue sites at the upper boundary are expected to vanish at the maximum strain strength, i.e., $J(1-\gamma \Delta u_1)=J(1-c_{max}y_{max})=0$, where the maximum $y$-coordinate is $y_{max}=\frac{3}{2}(L_y-1)+\frac{1}{2}$. Hence we have $c_{max}=1/y_{max}$, which we take as the scale of the strain parameter $c$ throughout the manuscript.

In the following discussions, we study the model in Eq.(1) under the above nonuniform unaxial strain using LSWT and stochastic series expansion (SSE) QMC method with directed loop updates~\cite{sandvik2002,syljuasen2003}. The SSE method expands the partition function in power series and the trace is written as a sum of diagonal matrix elements. The directed loop updates make the simulation very efficient~\cite{Bauer2011,fabien2005,pollet2004}. Our simulations are on a honeycomb nanoribbon with the total number of sites $N_s=L_x\times L_y$ with $L_x=20, L_y=200$ the linear sizes. The nanoribbon is periodic (open) along the $x(y)$ direction.  There are no approximations
causing systematic errors, and the discrete configuration space can be sampled without floating
point operations. The temperature is set to be $\beta=200$, which is low enough to obtain the ground-state properties.

\section{Magnon Landau levels in the linear spin wave theory}
Let us first investigate the physical properties of the strained model Eq.(1) by LSWT, where the spin operators are replaced by bosonic ones via Holstein-Primakoff (HP) transformation\cite{hptransformation}. The transformation on sublattice A (the spin is in the positive $z$-direction) is defined as
\begin{align}\label{eq2}
S^+_{i}&=\sqrt{2S}a_{i}, S^-_{i}=\sqrt{2S}a^{\dagger}_{i},\\ \nonumber
S^z_{i}&=S-a^{\dagger}_{i}a_{i}.
\end{align}
On sublattice B (the spin is in the negative $z$-direction), the spin operators are defined as
\begin{align}\label{eq3}
S^+_{i}&=\sqrt{2S}b^{\dagger}_{i},S^-_{i}= \sqrt{2S}b_{i},\\ \nonumber
S^z_{i}&=b^{\dagger}_{i} b_{i}-S.
\end{align}
Keeping only the bilinear terms, the bosonic tight binding Hamiltonian reads
\begin{align}
H=\sum_{\langle i j\rangle} J_{ij} S\left(a_{i} b_{j}+a_{i}^{\dagger} b_{j}^{\dagger}+a_{i}^{\dagger} a_{i}+b_{j}^{\dagger} b_{j}\right).
\end{align}
Performing a Fourier transformation in the $x$-direction and under the basis $X^{\dagger}_{k_x}=(a_{1,k_x}^\dagger, b_{1,k_x}, ...,a_{L_y,k_x}^\dagger, b_{L_y,k_x})$, the above Hamiltonian writes as $H=\sum_{k_x}X^{\dagger}_{k_x}M(k_x)X_{k_x}$, where $M(k_x)$ is a $2L_y\times 2L_y$ matrix.
By a standard Bogliubov transformation\cite{PhysRev.139.A450,xiao2009theory}, the matrix $M(k_x)$ becomes diagonal, and the magnon spectra are directly obtained.

\begin{figure}[htbp]
\centering \includegraphics[width=9.cm]{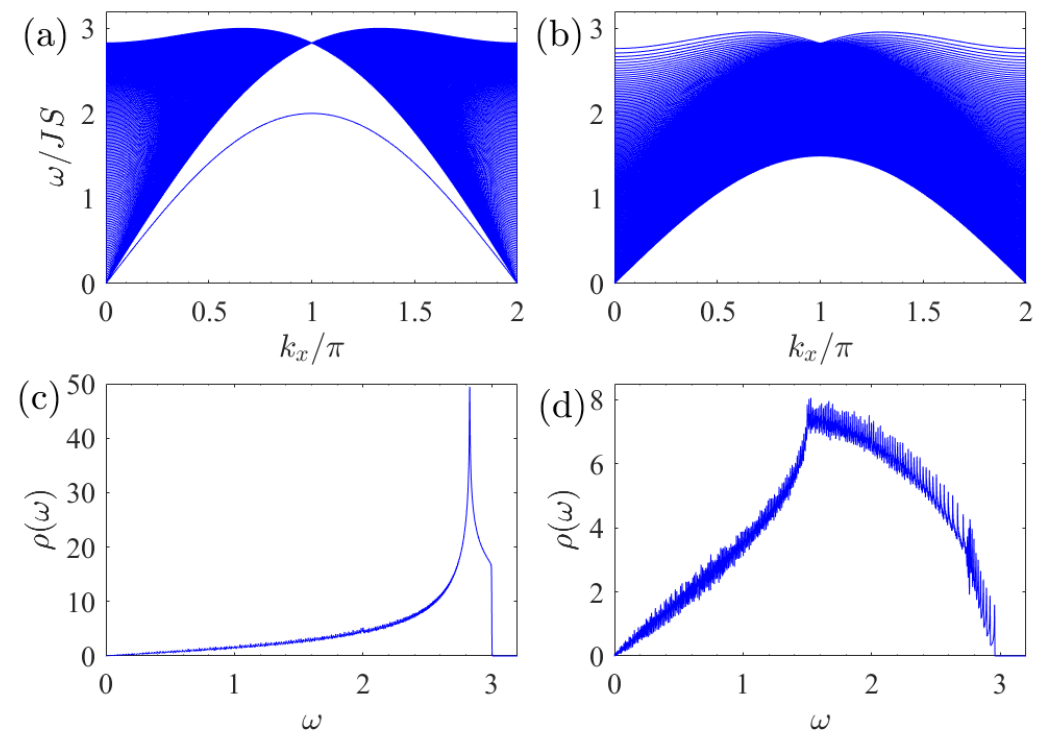} \caption{The magnon spectrum of the AF honeycomb nanoribbon: (a) without the strain; (b) in the presence of a non-uniform unaxial strain. (c) and (d) are the corresponding magnon density of states of (a) and (b), respectively. The strain strength in (b) and (d) is $c/c_{max}=1$.}
\label{fig2}
\end{figure}

The open boundaries are created by breaking the bonds connecting the outmost sites of the zigzag edges. As shown in Fig.\ref{fig2}(a), a new branch of modes associated with the boundaries appear below the bulk spectrum~\cite{2017Edge}. Their boundary nature is further revealed by the distribution of the corresponding wavefunctions, which is mainly localized near the boundaries. The density of states(DOS) is plotted in Fig.\ref{fig2}(c), which resembles that of itinerant electrons in graphene. As expected, the low-energy linear behavior in DOS is due to the linear dispersion of the magnon excitation in the antiferromagnets. Besides, the saddle point at $k_x=0$ leads to a Van Hove singularity in the magnon spectrum~\cite{VanHsingularity}.

\begin{figure}[htbp]
\centering \includegraphics[width=8.8cm]{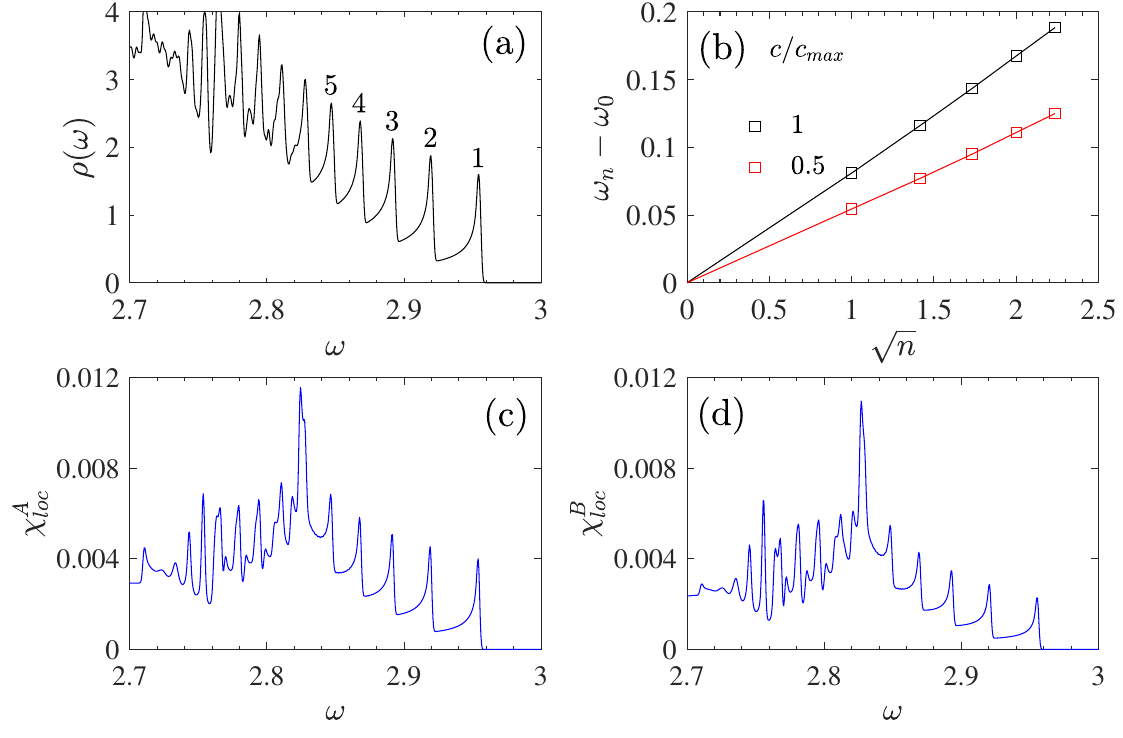} \caption{(a) The density of states near the upper end of the magnon spectrum. (b) The PLL energy
$\omega_n$ as a function of the square root of the level index $n$. The solid lines represent linear fitting of the data. Local susceptibility at the center of the unit cell on: (c) A sublattice and (d) B sublattice. The linear size is $L_x=20, L_y=200$. In (a),(c) and (d), the strain strength is $c/c_{max}=1$.}
\label{fig3}
\end{figure}

After the strain is applied, the degeneracy of the energy levels is removed, and the spectrum becomes much broader[see Fig.\ref{fig2}(b)]. This change is most evident at $k_x=\pi$, where all energy levels are originally degenerate in the absence of strain (see Appendix A). In particular, the magnon spectrum is flatted by the strain, and DOS exhibits oscillating behavior.
The appearance of sharp peaks in the magnon DOS should result from the flat levels, thus is a direct evidence of the formation of the magnon PLLs.
As shown in Fig.\ref{fig3}(a), the magnon PLLs appear from the upper end of the spectrum, which is in agreement with the recent studies on the Heisenberg model under a triaxial strain. However, by fitting the positions of the peaks, it is found that the PLL energy $\omega_n$ is proportional to the square root of the level index $n$, which is in great contrast to the equally spaced PLLs in honeycomb antiferromagnets under a triaxial strain. In addition, the scope increases when enhancing the applied strain. Except the appearing position, these properties are very similar to the Landau levels of Dirac fermions in graphene~\cite{RevModPhys.83.1193}(an analytical understanding is presented in Appendix B).

The appearance of the magnon PLLs can also be reflected in the local susceptibility, which is defined as\cite{PhysRevB.71.104427}
\begin{align}
\chi^{i}_{loc}(\omega)=\int_{-\infty}^{\infty} d t e^{i \omega t}\left\langle S_{i}^{x}(t) S_{i}^{x}+S_{i}^{y}(t) S_{i}^{y}\right\rangle.
\end{align}
In LSWT, the local susceptibility is formulated in terms of the $\delta$ functions peaked at the magnon eigenvalues, hence is equivalent to the density of states in characterizing the flat PLLs. More importantly, the local susceptibility can be exactly determined by numerical analytic continuation of the imaginary time spin correlations obtained by the QMC simulation. The local susceptibilities on $A$- and $B$-sublattice sites deep in the lattice are shown in Fig.\ref{fig3}(c) and (d). Indeed, both of them demonstrate sharp peaks at exactly the same positions with those in DOS, further confirming the formation of the magnon PLLs.

\section{The evolution of the AF order}

\begin{figure}[htbp]
\centering \includegraphics[width=9.cm]{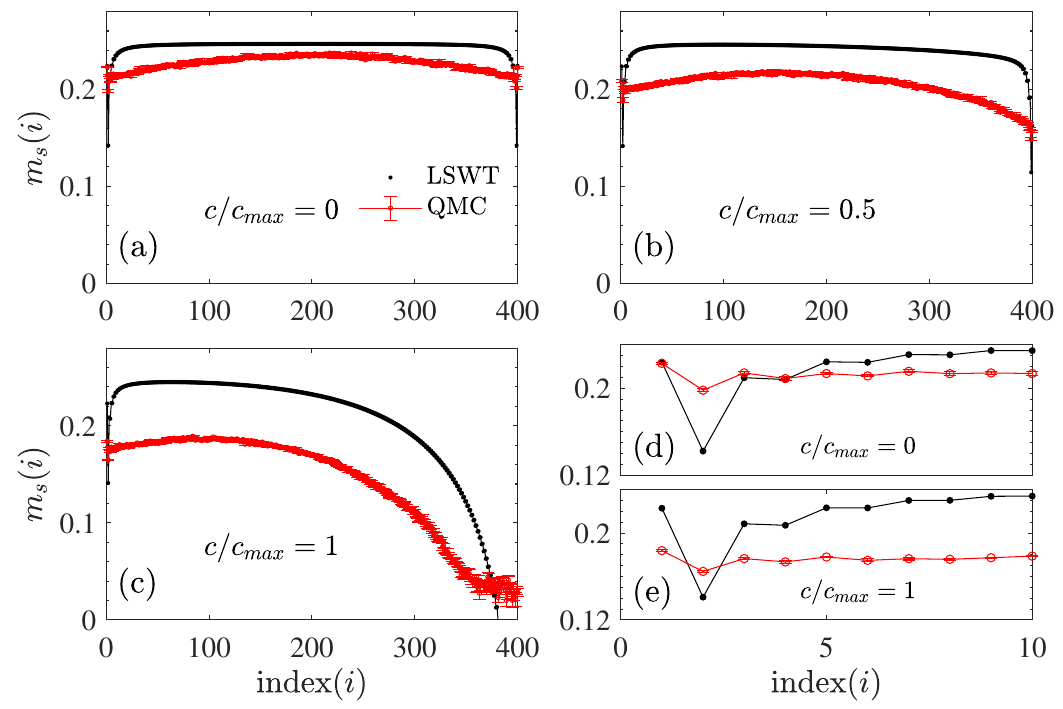} \caption{The distribution of the local magnetization obtained by LSWT and QMC simulations at the strain strength (a) $c/c_{max}=0$, (b) $0.5$, (c) $1$. (d) and (e) enlarge the curves near the lower boundary in (a) and (c), respectively. Here the index range is up to $2L_y=400$ (the same with the figures hereafter), which is due to the existence of the sublattice degree. }
\label{fig4}
\end{figure}

We next study how the AF order is affected by the strain. In LSWT, the existence of N{\'e}el order is identified by a finite local magnetization. Since the honeycomb nanoribbon is translation invariant in the $x$-direction, the local magnetization only varies within the unit cell which extends over the entire width of the ribbon (see Fig.1). Figure \ref{fig4} shows the local magnetization as a function of site index in the unit cell at several values of the strain strength. In the absence of strain, the N{\'e}el orders near the boundaries are perturbed, and the values gradually decrease as the sites approach the boundaries. Nevertheless, $m_s(i)$ is always finite, and becomes almost uniform away from the boundaries, implying the long-range AF order still preserves in the presence of open boundaries. It is noted that the local magnetization on the outmost sites of the boundary
is much larger than that of its nearby sites [see Fig.\ref{fig4}(d) and (e)]. The two outmost sites represent the two sublattices of the boundary zigzag chain. Although it is antiferromagnetic along the one-dimensional (1D) chain, the magnetic moments are unequal within the two-site unit cell, resulting in a net ferromagnetic moment.  Hence a
ferrimagnetic order is formed along the zigzag boundary,
which has also been revealed in the Hubbard model on
honeycomb nanoribbons~\cite{PhysRevB.68.035432,PhysRevLett.112.046601,PhysRevB.87.155441,PhysRevB.81.115416,PhysRevX.4.021042}.

After the strain is applied, the value of the local magnetization monotonically decreases with the strain strength. Since the exchange coupling is gradually reduced in the $y$-direction, the magnetization is more affected on the sites father away from the lower boundary.
In particular, at large enough strain strength and near the upper boundary the local magnetization decreases rapidly and becomes negative at a critical position, implying the AF order vanishes hereafter. This behavior is due to the exchange couplings of the vertical bonds become negligibly small near the upper boundary, and the system can be regarded as a collection of isolated 1D Heisenberg chains, resulting the breakdown of 2D AF order there.
In contrast, the region near the lower boundary is less affected since the exchange couplings here are least modified.
While the LSWT can qualitatively demonstrate the evolution of the AF order with the strain, the exact results should be obtained by the unbiased QMC simulations.

In QMC simulations, the local value of the magnetization is given by $m_s^{qmc}(i)$, defined as\cite{PhysRevLett.90.177205}
\begin{align}
m_{s}^{qmc}(i)=\sqrt{\frac{3}{N} \sum_{j=1}^{N}\textrm{sgn}(i,j)\left\langle S_{i}^{z} S_{j}^{z}\right\rangle},
\end{align}
where the sum is over all lattice sites $j$, and $\textrm{sgn}(i,j)=1(-1)$ if $i,j$ belong to the same (opposite) sublattice. Figure 4 plots the values of $m_s^{qmc}(i)$ at the same strain strengths as those in LSWT.
The QMC values are smaller than the LSWT ones. Besides, the difference between the values from the two approaches increases as the strain is strengthened. In the absence of strain, the QMC curve slowly increases and gets a maximum at the central point. In contrast, the LSWT one is nearly flat in most of the bulk region. Here it is noted that the QMC and LSWT results are only slightly different, and most of the values from the two approaches have less than a $10$ percent difference. This implies the linear approximation in the HP transformation is pretty accurate, which has also been found in the existing literature\cite{dalla2015fractional,PhysRevX.7.041072}. For the strain strength $c/c_{max}=1$, a clear transition is visible in the QMC curve near the upper boundary.

\begin{figure}[htbp]
\centering \includegraphics[width=9.cm]{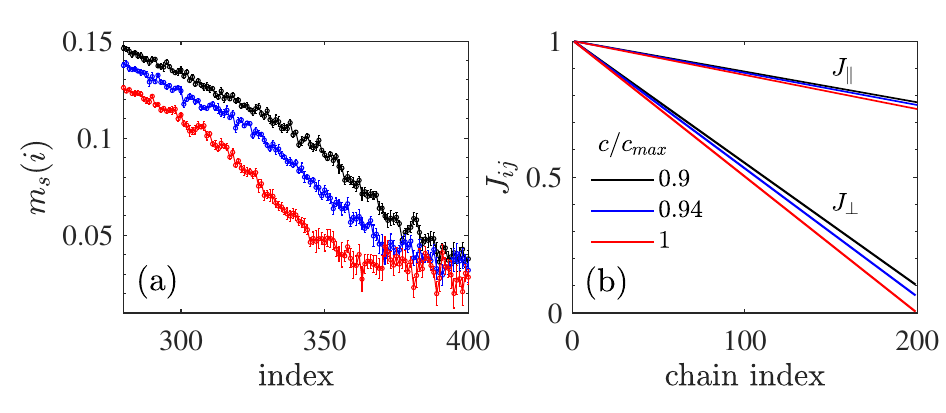} \caption{(a) The local magnetization obtained by QMC near the critical position for various strain strengths. (b) The paralell and transverse exchange couplings as a function of the zigzag chain index. }
\label{fig5}
\end{figure}

Figure \ref{fig5}(a) plots the local magnetization obtained by QMC near the critical position for various strain strengths. It shows the crossover from 2D to 1D behavior is continuous, which may be due to that the exchange coupling varies smoothly all the way down to very small value with the coordinate $y$ [see Figure \ref{fig5}(b)]. We can not determine the exact critical strain after which there appears such a transition. Nevertheless, since the transition has already become indistinguishable at $c/c_{max}=0.9$, the critical value should be pretty large.
While such a transition is also predicted by LSWT with $m_s(i)=0$, the QMC transition happens a bit deeper in the ribbon than the LSWT one.
These results imply that although the quantum fluctuation is omitted, LSWT can still give qualitatively correct evolution of the AF order.

\begin{figure}[htbp]
\centering \includegraphics[width=8.8cm]{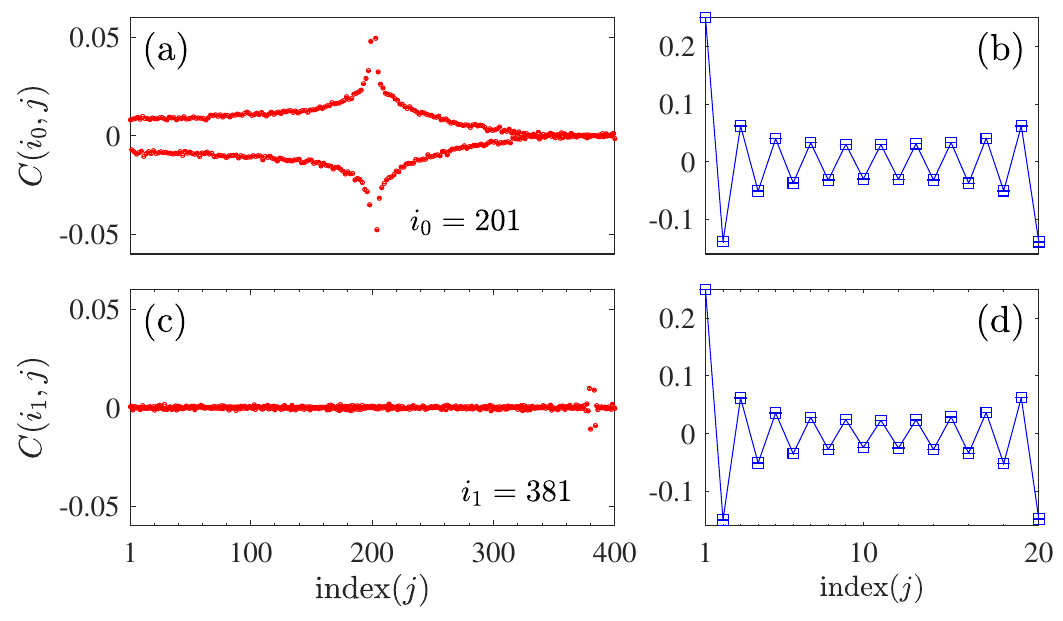} \caption{The spin correlation with the reference point fixed at the middle point $i_0$ of the unit cell: (a) $j$ varies within the unit cell, (b) $j$ is on the same zigzag chain with $i_0$ (here $i_0$ corresponds to the $j=1$ site). The reference site is changed to a near-boundary site $i_1$ in (c) and (d), which are the corresponding plots of (a) and (b), respectively.}
\label{fig6}
\end{figure}

How the magnetic property is affected by the strain can also be demonstrated by the spin correlation $C(i,j)=\langle S_i^zS_j^z\rangle$. Figure \ref{fig6} plots the spin correlation between two sites within the super unit cell at $c/c_{max}=1$ for the Heisenberg Hamiltonian. When the reference point $i_0$ is in the middle of the unit cell, $C(i_0,j)$ is always finite for $j<i_0$ ($j$ is located in the lower part), but it gradually decreases in the upper part and becomes nearly zero from a critical position. In contrast, for a reference point $i_1$ near the upper boundary, $C(i_1,j)$ reduces to zero quickly as $j$ goes away from $i_1$. We also plot $C(i,j)$ with both $i,j$ on the same zigzag chain, which keeps finite even for the largest distance. These results are consistent with the occurrence of a crossover from 2D to 1D magnetic properties at the critical position. Moreover, it is noted in Fig.5(a) that the curve begins to decrease in a slower way after the critical position. This can be understood in terms of the spin correlations. Since the vertical exchange coupling has become negligibly weak near the upper boundary, the spin correlation in this direction is nearly zero. In contrast, the spin correlations along the zigzag chain are still considerably large, which actually dominates the local magnetization. The parallel spin correlations vary slowly with $y$ here, and so does the local magnetization. Due to the contribution from the parallel spin correlations, the local magnetization still has a finite small value.

\section{The strained $XY$ antiferromagnetic honycomb nanoribbon}

We next consider the spin-$\frac{1}{2}$ $XY$ AF Hamiltonian described by
\begin{align}
H_{XY}=J\sum_{\langle ij\rangle}(S_i^xS_j^x+S_i^yS_j^y).
\end{align}
By a rotation of the coordinate system, the above model becomes the $XZ$ Hamiltonian\cite{Joannopoulos1987}
\begin{align}
H_{XZ}=&J\sum_{\langle ij\rangle}(S_i^xS_j^x+S_i^zS_j^z),  \\ \nonumber
=&J\sum_{\langle ij\rangle}(S_i^zS_j^z+\frac{1}{4}\sum_{\mu,\nu=\pm}S_i^{\mu}S_j^{\nu}).
\end{align}
Under a non-uniform unaxial strain, the same modulation of the exchange coupling with that in Eq.(2) can be made, and the application of LSWT is straightforward.

\begin{figure}[htbp]
\centering \includegraphics[width=8.8cm]{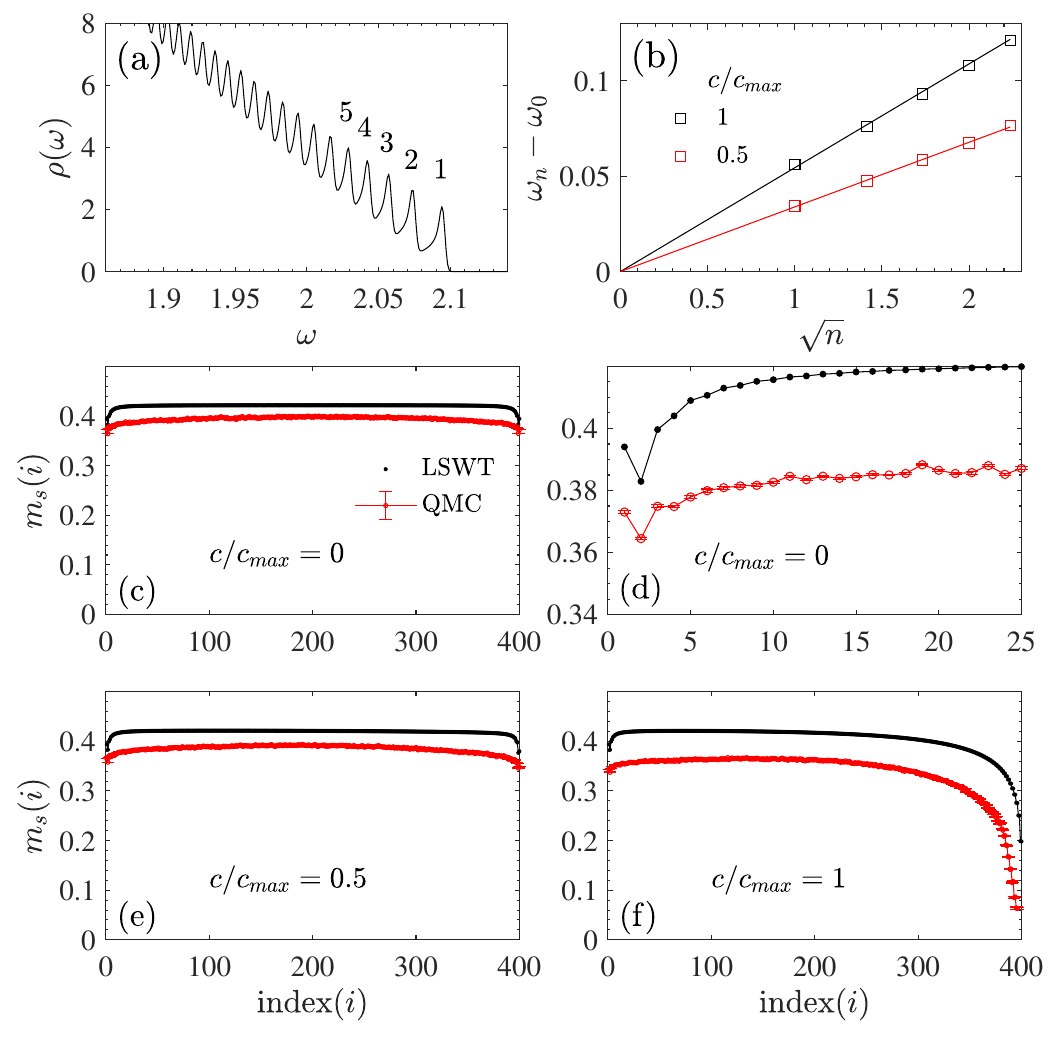} \caption{ (a) The magnon density of states near the upper end of the spectrum of the strained $XY$ Hamiltonian. (b) The PLL energy
$\omega_n$ as a function of the square root of the level index $n$. The distribution of the local magnetization obtained
by LSWT and QMC simulations at the strain strength $c/c_{max}$: (c)
$0$; (e) $0.5$; (f) $1$. (d) enlarges the curves of (c) near
the lower boundary. In (a), the strain strength is $c/c_{max}=1$.
}
\label{fig7}
\end{figure}

Figure \ref{fig7}(a) plots the magnon density of states. Under PMF induced by the strain, sharp peaks appear from the upper end of the spectrum, marking the formation of PLLs. Besides, the energies of PLLs are proportion to $\sqrt{nc}$ [$n$ is the level index, and the PMF magnitude is proportional to the strain strength $c$, see Fig.\ref{fig7}(b)], which is very similar to the results from the strained Heisenberg model. However, such properties are in great contrast to the situation under a triaxial strain, where PLLs appear from the middle of the spectrum and the peaks follow the relations $\propto n^{\frac{1}{3}}, n^{\frac{2}{3}}$~\cite{sun2021quantum}.

We then investigate how the strain affects the AF order. Similarly, the local magnetization decreases
monotonically in the y-direction near the uppper boundary in the presence of strain.
Compared to the Heisenberg case, the values of the local magnetization is much larger at the same condition. Besides, as shown in Fig.\ref{fig7}(f), even at the largest strain strength when the $y$-direction bonds near the upper boundary is considerably weak, the local magnetization always keeps finite, suggesting the long-range AF order preserves in the whole system. This implies the $XY$ Hamiltonian is more robust to the modulation of the exchange couplings induced by the strain~\cite{guo2021quantum}. Qualitatively, the reason is that there are three (two) spin components in the Heisenberg ($XY$) case, thus the quantum fluctuation is much stronger in the Heisenberg model than that in the $XY$ one.
The related quantity defined in Eq.(9) is also calculated using the QMC methods. While the QMC results are qualitatively consistent with those from LSWT, QMC gives relatively smaller values at the same strain strengths.

\section{Conclusions}

The magnon PLLs and the evolution of the AF order in the strained AF honeycomb nanoribbons are studied using LSWT and QMC simulations. After the strain is applied, the magnon PLLs are formed from the upper end of the spectrum, and their level spacings are proportional to the square root of the level index. Since the exchange couplings are linearly weakened by the unaxial strain, the local magnetization
decreases monotonically in the $y$-direction. Specifically, at large enough strain strength, the $y$-direction bonds near the upper boundary become negligibly weak such that the system there is decoupled into isolated zigzag chains, exhibiting one-dimensional antiferromagnetic property.
The $XY$ Hamiltonian under the same kind of strain demonstrates similar properties except that the AF order is more robust than the Heisenberg case. The behavior of the $XY$ case is in great contrast to that under a
triaxial strain, where the PLLs appear from the middle of
the spectrum and the peaks follow a third-root relation.

In the past several years, significant progress has been achieved in the field of 2D quantum magnetic materials\cite{Yi-LinZhang:27501,burch2018magnetism,gibertini2019magnetic,PhysRevX.10.011062}. 2D magnetic order has been observed in various magnetic van der Waals materials, and most of them form with the magnetic elements in a honeycomb lattice\cite{material_1,Wang_2016,gong2017discovery,bonilla2018strong,material_2}. All the 2D magnetism can, in principle, be described by three fundamental models: Ising, $XY$, or Heisenberg. Strain engineering, as an important approach to control and manipulate magnetic states, has been widely adopted in the research of 2D magnetic materials\cite{PhysRevMaterials.4.094004,Mukherjee_2019,strain1,Rold_n_2015}. Thus the new physical phenomena induced by the engineered strain would definitely be interesting to the related experiments. Moreover, the present study will contribute to the theoretical understanding of the behavior of the neutral quasi-particles in pseudo-magnetic fields, and propose an alternative routine to manipulate magnons, which may have potential applications in designing new devices of magnon spintronics\cite{chumak2015magnon,PhysRevX.9.011026}.

\section*{Acknowledgments}
The authors thank Tianyu Liu, Yancheng Wang, Wen Yang, Chenyue Wen, Xingchuan Zhu for helpful discussions. J.S and H.G. acknowledges support from the NSFC grant Nos.~11774019 and 12074022, the NSAF grant in NSFC with grant No. U1930402, the Fundamental Research
Funds for the Central Universities and the HPC resources
at Beihang University.
S.F. is supported by the National Key Research and Development Program of China under Grant No. 2016YFA0300304, and NSFC under Grant Nos. 11974051 and 11734002.

\appendix

\renewcommand{\thefigure}{A\arabic{figure}}

\setcounter{figure}{0}

\section{The analytical solution of the magnon eigenvalues at $k_x=\pi$ in Fig.\ref{fig2}(a)}

In the absence of strain, the Hamiltonian matrix of Eq.(7) in the momentum space reads as
\begin{align}\label{aeq1}
M(k_x)=JS\left(
         \begin{array}{ccccccc}
           2 & \gamma_{k_x} & 0 & 0 & 0 & 0 & \cdots \\
           \gamma^*_{k_x} & 3 & 1 & 0 & 0 & 0 & 0 \\
           0 & 1 & 3 & \gamma^*_{kx} & 0 & 0 & 0 \\
           0 & 0 & \gamma_{kx} & 3 & 1 & 0 & 0 \\
           0 & 0 & 0 & 1 & 3 & \gamma_{kx} & 0 \\
           0 & 0 & 0 & 0 & \gamma^*_{kx} & 3 & \cdots \\
           \vdots & 0 & 0 & 0 & 0 & \vdots & \ddots \\
         \end{array}
       \right),
\end{align}
where $\gamma_{k_x}=1+e^{-ik_x}$. At $k_x=\pi$, we have $\gamma_{k_x=\pi}=0$, thus $M(k_x)$ is block diagonal. There are two eigenvalues: $2J, 2\sqrt{2}J$, among which the value $2\sqrt{2}J$ has a large degeneracy. The strain breaks the degeneracy, and the spectrum is broadened at $k_x=\pi$, as shown in Fig.\ref{fig2}(b).

\section{The analytical treatment based on the effective Hamiltonian near the Dirac point}
Here we present the analytical treatment of magnon pseudo-Landau levels in the stained quantum antiferromagnetic Heisenberg model based on the effective Hamiltonian near the Dirac point.

\subsection{The low-energy effective Hamiltonian near the Dirac point}
After the spin operators are replaced by bosonic ones via Holstein-Primakoff transformation, we obtain the following bosonic tight-binding Hamiltonian,
\begin{align}\label{beq1}
\nonumber
  H&=J_1S\sum_{i,j}\left(a_i^\dagger b_j^\dagger+a_i b_j+a_i^\dagger a_i+b_j^\dagger b_j\right)\\ \nonumber
  &+J_2S\sum_{i,j}\left(a_i^\dagger b_j^\dagger+a_i b_j+a_i^\dagger a_i+b_j^\dagger b_j\right)\\
  &+J_3S\sum_{i,j}\left(a_i^\dagger b_j^\dagger+a_i b_j+a_i^\dagger a_i+b_j^\dagger b_j\right).
\end{align}
In the momentum space, the Hamiltonian becomes $H=\sum_{\bf k}\hat{\Psi}_{\bf k}^\dagger h({\bf k})\hat{\Psi}_{\bf k}$ with the basis $\hat{\Psi}_{\bf k}^\dagger=\left(a_{\bf k}^\dagger\ \ b_{\bf k}\right)$, and
\begin{align}
\nonumber
&h(\bm{k})=d_x(k)\sigma_x+d_y(k)\sigma_y+\left(J_1+2J_2\right)S\sigma_0, \\ \nonumber
&d_{x}(\boldsymbol{k})=J_{1}S \cos k_{y}+2 J_{2}S \cos \frac{\sqrt{3} k_{x}}{2} \cos \frac{k_{y}}{2}, \\ \nonumber
&d_{y}(\boldsymbol{k})=-J_{1}S \sin k_{y}+2 J_{2}S \cos \frac{\sqrt{3} k_{x}}{2} \sin \frac{k_{y}}{2},
\end{align}
where $\sigma_{x,y}$ are the Pauli matrices; $\sigma_0$ is the identity matrix; $J_n=J(1-\gamma\Delta u_n)$ with $\Delta u_{1}=\epsilon_{y y}, \Delta u_{2}=\Delta u_{3}=\epsilon_{y y} / 4$.
Writing the momentum near the Dirac point $\bm{K}=\left(\frac{4\pi}{3\sqrt3},0\right)$ as $\bm k=\bm K +\bm q$, and expanding $d_{y}(\boldsymbol{k}),d_{x}(\boldsymbol{k})$ to linear order of ${\bf q}$, the resulting Hamiltonian is,
\begin{align}\label{beq2}
\nonumber
&h(\bm q)=-\frac32JS\left[\left(1-\frac{1}{4}\epsilon_{yy}\right)q_x+\frac12\epsilon_{yy}\right]\sigma_x \\ \nonumber
&-\frac32JS\left[\left(1-\frac{3}{4}\epsilon_{yy}\right)q_y+\frac{q_xq_y}{2}\left(1-\frac14\epsilon_{yy}\right)\right]\sigma_y \\
&+JS\left(3-\frac32\epsilon_{yy}\right)\sigma_0.
\end{align}

The strain tensor is expected to generate a pseudo-magnetic field, and the vector potential is,
\begin{align}\label{beq3}
  \vec{A}=\frac{\gamma}{2}\left(
                            \begin{array}{c}
                              \epsilon_{xx}-\epsilon_{yy}  \\
                              -2\epsilon_{xy}  \\
                            \end{array}
                          \right).
\end{align}
We choose $\epsilon_{yy}=\frac{c}{\gamma}y$ to get a homogeneous field $\vec{B}=\frac{1}{2}c\hat{z}$. Introducing $p=\frac12-\frac{q_x}4, s=1+\frac{q_x}2, r=\frac34+\frac{q_x}{8}$, and changing $y\rightarrow y+\frac{s}{rc}$, $k_y\rightarrow -i\partial_y$, we get the effective Hamiltonian near the Dirac point,
\begin{align}\label{beq4}
\nonumber
h(q_x)=&-\frac32JS\left\{\left[q_x+pc\left(y+\frac{s}{rc}\right)\right]\sigma_x+ircy\partial_y\sigma_y\right\} \\
&+3JS(1-\frac{1}{2}cy)\sigma_0.
\end{align}

\subsection{A solvable case without the $y$-dependent term before $\sigma_0$}
We first consider a specific solvable case: the $y$-dependent term before $\sigma_0$ is dropped artificially. The Hamiltonian reads as follows,
\begin{align}\label{beq5}
h(q_x)=-\frac32JS\left\{\left[q_x+pc\left(y+\frac{s}{rc}\right)\right]\sigma_x+ircy\partial_y\sigma_y\right\}+3JS\sigma_0
\end{align}
We consider the following eigenvalue problem,
\begin{equation}\label{beq6}
  \tau_zh(q_x)\left(
                    \begin{array}{c}
                      \phi_A(y) \\
                      \phi_B(y) \\
                    \end{array}
                  \right)=E\left(
                    \begin{array}{c}
                      \phi_A(y) \\
                      \phi_B(y) \\
                    \end{array}
                  \right),
\end{equation}
where $\tau_z$ is the Pauli matrix.
Expanding the above matrix-vector multiplication, two first order differential equations are obtained,
\begin{align}\label{beq7andbeq8}
  -\frac32&\left[q_x+pc\left(y+\frac{s}{rc}\right)+rc\left(y\partial_y+\frac12\right)\right]\phi_B=\left(\varepsilon-3\right)\phi_A,\\
  \frac32&\left[q_x+pc\left(y+\frac{s}{rc}\right)-rc\left(y\partial_y+\frac12\right)\right]\phi_A=\left(\varepsilon+3\right)\phi_B,
\end{align}
where $\varepsilon=\frac{E}{\left(JS\right)}$.

\begin{figure}[htbp]
\centering \includegraphics[width=8.8cm]{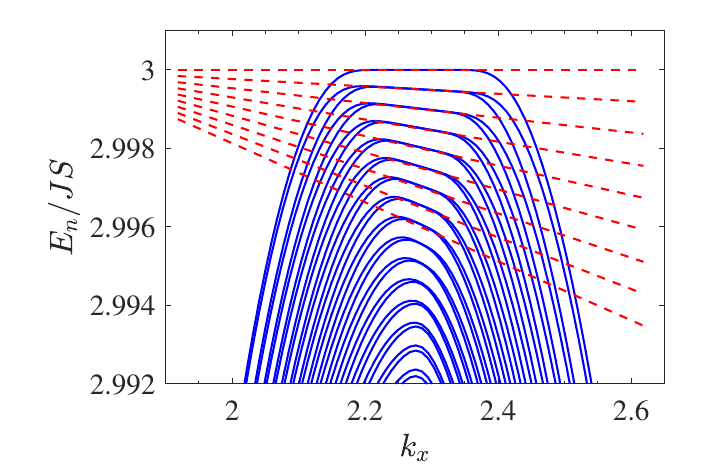} \caption{Comparison bewteen the analytical solution Eq.(\ref{beq12}) and the dispersion
obtained by numerically diagonalizing the Hamiltonian matrix near one of the Dirac points. Here the $y$-dependent term
before $\sigma_0$ in Eq.(\ref{beq4}) is dropped artificially. The strain strength is $c/c_{max}=0.5$. The linear size used in the numerical diagonalization is $L_y=200$.
}
\label{afig1}
\end{figure}

Eliminating $\phi_A$ by substituting Eq.(B7) into Eq.(B8), we obtain a second order ordinary differential equation with variable coefficient,
\begin{align}\label{beq9}
  y^2\phi_B^{\prime\prime}+2y\phi_B^\prime-\left[\frac{p^2}{r^2}y^2+\eta y+\frac{\Delta}{c^2r^4}-\frac{1}{4}\right]\phi_B=0,
\end{align}
where $\Delta=(q_xr+ps)^2-4r^2+r^2\left(\frac23\varepsilon\right)^2$ and $\eta=\frac{p}{cr^3}(2q_xr+2ps-cr^2)$.
We first examine the asymptotic form of the solution. As $y\rightarrow -\infty$, the $y^2$ term dominates, so $\phi^{''}_B-\frac{p^2}{r^2}\phi_B=0$. The general solution is $\phi_B=Ae^{-\frac{p}{r}y}+Be^{\frac{p}{r}y}$. Since $e^{-\frac{p}{r}y}$ diverges at $y\rightarrow -\infty$, $\phi_B\sim e^{\frac{p}{r}y}$. Similarly, at $y=0$, $\phi_B\sim e^{-\frac{1}{2}+\frac{\sqrt{\Delta}}{cr^2} }$. Taking these asymptotic behavior into consideration, we can write the eigenfunction as $\phi_B=e^{\frac{p}{r}y} e^{-\frac{1}{2}+\frac{\sqrt{\Delta}}{cr^2} }u(y)$, so that the differential equation can be simplified. In terms of $u(y)$, Eq.(\ref{beq9}) becomes,
\begin{align}\label{beq10}
&yu^{''}+(1+\frac{2\sqrt{\Delta}}{cr^2}+\frac{2p}{r}y)u' \\ \nonumber
&+\frac{2p}{r}(1+\frac{\sqrt{\Delta}}{cr^2}-\frac{q_xr+ps}{cr^2})u=0.
\end{align}
Introducing $\gamma=1+\frac{\sqrt{\Delta}}{cr^2}$, $\alpha=1+\frac{\sqrt{\Delta}}{cr^2}-\frac{q_xr+ps}{cr^2}$, and $z=-\frac{2p}{r}y$, we arrive in the confluent hypergeometric equation, $zu''(z)+(\gamma-z)u'(z)-\alpha u(z)=0$. The above differential equation has a regular singularity at $z=0$, and can be solved by the series expansion method. One solution is,
\begin{align}\label{beq11}
  u(z)=1+\frac{\alpha}{\gamma}\frac{z}{1!}+\frac{\alpha(\alpha+1)}{\gamma(\gamma+1)}\frac{z^2}{2!}+\cdots, \gamma\neq 0,-1,-2,\cdots.
\end{align}
To make $u(z)$ a polynomial so that finite, $\alpha$ should be $0$ or a negative integer, i.e., $\alpha=-n, n=0,1,2,\cdots$. Then we get the following expression for the eigenenergy,
\begin{align}\label{beq12}
  E_n=3JS\sqrt{1-\frac{2+3q_x}{8}nc}.
\end{align}
In the limit of small $c$, we can approximate the eigenenergy as $E_n\approx 3JS(1-\frac{2+3q_x}{16}nc)$, which implies the pseudo-Landau levels are equally-spaced with the level index $n$. We compare the analytical solution with the dispersion
obtained by numerically diagonalizing the Hamiltonian matrix. As shown in Fig.\ref{afig1} the results are in very good consistence near the Dirac point, which further verifies our calculations in the paper.

\begin{widetext}

\subsection{Attempt to solve the full effective Hamiltonian}

Expanding the eigenvalue problem of the full effective Hamiltonian, we obtain two first-order differential equation,
\begin{align}\label{beq13andbeq14}
  -\frac32 &\left[q_x+pc\left(y+\frac{s}{rc}\right)+rc\left(y\partial_y+\frac12\right)\right]\phi_B=
  \left\{\varepsilon-\left[3-\frac32c\left(y+\frac{s}{rc}\right)\right]\right\}\phi_A,\\
  \frac32 &\left[q_x+pc\left(y+\frac{s}{rc}\right)-rc\left(y\partial_y+\frac12\right)\right]\phi_A=
  \left\{\varepsilon+\left[3-\frac32c\left(y+\frac{s}{rc}\right)\right]\right\}\phi_B.
\end{align}
Defining $\alpha(y)=\frac32c\left(y+\frac{s}{rc}\right)$ and eliminating $\phi_A$ in Eq.(B14), we get the following second-order differential equation of $\phi_B$,
\begin{align}\label{beq15}
\nonumber
 &\left\{\left[y^2+\frac{\alpha}{(\varepsilon-3)} y^2\right]\phi_B^{\prime\prime}+\left[2y+\frac{2\alpha y-y^2\alpha^\prime(y)}{(\varepsilon-3)} \right]\phi_B^\prime\right\}\\ \nonumber
 -\phi_B&\left\{\frac{y^2}{r^2} \left[p^2+\frac{p( p\alpha+  r\alpha^\prime)}{(\varepsilon-3)}\right]+y \left[\frac{p\left(2ps+2q_x r-cr^2\right)}{cr^3}+\alpha\frac{p\left(2ps+2q_x r-cr^2\right)}{cr^3(\varepsilon-3)}+\alpha^\prime\frac{(cr^2+2ps+2q_xr)}{2cr^2(\varepsilon-3)}\right]\right\}\\ \nonumber
 &-\phi_B\left\{\frac{(ps+q_xr)^2}{c^2r^4}-\frac14+\alpha\frac{4(ps+q_xr)^2-c^2 r^4}{4c^2r^4(\varepsilon-3)}\right\}\\
 &=-\frac{4}{9c^2r^2}\left\{(\varepsilon^2-9)+\frac{\left[(\varepsilon+3)-\alpha\right]}{(\varepsilon-3)}\left[2\alpha(\varepsilon-3)+\alpha^2\right]-\alpha(\varepsilon-3)\right\}\phi_B.
\end{align}
If $\alpha=0$ and $\alpha^{\prime}=0$ is set, the above equation reduces to Eq.(\ref{beq9}). Due to the presence of the $y$-dependent term
before $\sigma_0$, the differential equation becomes much more complex.
Defining $b=\frac{3}{2}c/(\varepsilon-3+\frac{3s}{2r})$, Eq.(\ref{beq15}) becomes,
\begin{align}\label{beq16}
 &\left(y^2+by^3\right)\phi_B^{\prime\prime}+\left(2y+by^2\right)\phi_B^\prime-A_0\phi_B,
\end{align}
with
\begin{align}\label{}
\nonumber
A_0&=b\frac{\left(p^2-1\right)}{r^2}y^3+\left\{\frac{2 b \left[p(ps+q_xr)-(s-2r)\right]}{c r^3}+\frac{p^2-1}{r^2}\right\}y^2\\ \nonumber
&+\left(\frac{4 b \varepsilon^2}{9 c^2r^2}+\frac{b (p s+q_x r)^2-b(s-2r)^2}{c^2 r^4}+\frac{b \left(p s+q_x r\right)}{cr^2}+\frac{2[p(ps+q_xr)-(s-2r)]}{cr^3}+\frac{b}{4}-\frac{p}{r}\right)y\\ \nonumber
&+\frac{4 \varepsilon^2}{9 c^2r^2}+\frac{(p s+q_x r)^2-(s-2r)^2}{c^2 r^4}-\frac{1}{4}.
\end{align}

We have tried to decouple the solution using the asymptotic forms at $y=0$ and $y\rightarrow -\infty$. However the resulting differential equation does not fit into any standard one. At present, it is still unclear for us whether the differential equation Eq.(\ref{beq16}) has an analytical solution.

\end{widetext}

\bibliography{ref}% Produces the bibliography via BibTeX.

\begin{thebibliography}{66}
\expandafter\ifx\csname natexlab\endcsname\relax\def\natexlab#1{#1}\fi
\expandafter\ifx\csname bibnamefont\endcsname\relax
  \def\bibnamefont#1{#1}\fi
\expandafter\ifx\csname bibfnamefont\endcsname\relax
  \def\bibfnamefont#1{#1}\fi
\expandafter\ifx\csname citenamefont\endcsname\relax
  \def\citenamefont#1{#1}\fi
\expandafter\ifx\csname url\endcsname\relax
  \def\url#1{\texttt{#1}}\fi
\expandafter\ifx\csname urlprefix\endcsname\relax\def\urlprefix{URL }\fi
\providecommand{\bibinfo}[2]{#2}
\providecommand{\eprint}[2][]{\url{#2}}

\bibitem[{\citenamefont{Pereira and
  Castro~Neto}(2009)}]{PhysRevLett.103.046801}
\bibinfo{author}{\bibfnamefont{V.~M.} \bibnamefont{Pereira}} \bibnamefont{and}
  \bibinfo{author}{\bibfnamefont{A.~H.} \bibnamefont{Castro~Neto}},
  \bibinfo{journal}{Phys. Rev. Lett.} \textbf{\bibinfo{volume}{103}},
  \bibinfo{pages}{046801} (\bibinfo{year}{2009}),
  \urlprefix\url{https://link.aps.org/doi/10.1103/PhysRevLett.103.046801}.

\bibitem[{\citenamefont{Guinea et~al.}(2010{\natexlab{a}})\citenamefont{Guinea,
  Katsnelson, and Geim}}]{Guinea2010Energy}
\bibinfo{author}{\bibfnamefont{F.}~\bibnamefont{Guinea}},
  \bibinfo{author}{\bibfnamefont{M.}~\bibnamefont{Katsnelson}},
  \bibnamefont{and} \bibinfo{author}{\bibfnamefont{A.}~\bibnamefont{Geim}},
  \bibinfo{journal}{Nature Physics} \textbf{\bibinfo{volume}{6}},
  \bibinfo{pages}{30} (\bibinfo{year}{2010}{\natexlab{a}}),
  \urlprefix\url{https://doi.org/10.1038/nphys1420}.

\bibitem[{\citenamefont{Yang et~al.}(2010)\citenamefont{Yang, Jiang, Liu, and
  Liu}}]{2010Strain}
\bibinfo{author}{\bibfnamefont{L.}~\bibnamefont{Yang}},
  \bibinfo{author}{\bibfnamefont{X.}~\bibnamefont{Jiang}},
  \bibinfo{author}{\bibfnamefont{Z.}~\bibnamefont{Liu}}, \bibnamefont{and}
  \bibinfo{author}{\bibfnamefont{Z.}~\bibnamefont{Liu}}, \bibinfo{journal}{Nano
  Research} \textbf{\bibinfo{volume}{3}}, \bibinfo{pages}{545}
  (\bibinfo{year}{2010}).

\bibitem[{\citenamefont{Amorim et~al.}(2016)\citenamefont{Amorim, Cortijo, {de
  Juan}, Grushin, Guinea, Gutiérrez-Rubio, Ochoa, Parente, Roldán, San-Jose
  et~al.}}]{AMORIM20161}
\bibinfo{author}{\bibfnamefont{B.}~\bibnamefont{Amorim}},
  \bibinfo{author}{\bibfnamefont{A.}~\bibnamefont{Cortijo}},
  \bibinfo{author}{\bibfnamefont{F.}~\bibnamefont{{de Juan}}},
  \bibinfo{author}{\bibfnamefont{A.}~\bibnamefont{Grushin}},
  \bibinfo{author}{\bibfnamefont{F.}~\bibnamefont{Guinea}},
  \bibinfo{author}{\bibfnamefont{A.}~\bibnamefont{Gutiérrez-Rubio}},
  \bibinfo{author}{\bibfnamefont{H.}~\bibnamefont{Ochoa}},
  \bibinfo{author}{\bibfnamefont{V.}~\bibnamefont{Parente}},
  \bibinfo{author}{\bibfnamefont{R.}~\bibnamefont{Roldán}},
  \bibinfo{author}{\bibfnamefont{P.}~\bibnamefont{San-Jose}},
  \bibnamefont{et~al.}, \bibinfo{journal}{Physics Reports}
  \textbf{\bibinfo{volume}{617}}, \bibinfo{pages}{1} (\bibinfo{year}{2016}),
  ISSN \bibinfo{issn}{0370-1573}, \bibinfo{note}{novel effects of strains in
  graphene and other two dimensional materials},
  \urlprefix\url{https://www.sciencedirect.com/science/article/pii/S0370157315005402}.

\bibitem[{\citenamefont{de~Juan et~al.}(2013)\citenamefont{de~Juan, Ma\~nes,
  and Vozmediano}}]{PhysRevB.87.165131}
\bibinfo{author}{\bibfnamefont{F.}~\bibnamefont{de~Juan}},
  \bibinfo{author}{\bibfnamefont{J.~L.} \bibnamefont{Ma\~nes}},
  \bibnamefont{and} \bibinfo{author}{\bibfnamefont{M.~A.~H.}
  \bibnamefont{Vozmediano}}, \bibinfo{journal}{Phys. Rev. B}
  \textbf{\bibinfo{volume}{87}}, \bibinfo{pages}{165131}
  (\bibinfo{year}{2013}),
  \urlprefix\url{https://link.aps.org/doi/10.1103/PhysRevB.87.165131}.

\bibitem[{\citenamefont{Castro~Neto et~al.}(2009)\citenamefont{Castro~Neto,
  Guinea, Peres, Novoselov, and Geim}}]{RevModPhys.81.109}
\bibinfo{author}{\bibfnamefont{A.~H.} \bibnamefont{Castro~Neto}},
  \bibinfo{author}{\bibfnamefont{F.}~\bibnamefont{Guinea}},
  \bibinfo{author}{\bibfnamefont{N.~M.~R.} \bibnamefont{Peres}},
  \bibinfo{author}{\bibfnamefont{K.~S.} \bibnamefont{Novoselov}},
  \bibnamefont{and} \bibinfo{author}{\bibfnamefont{A.~K.} \bibnamefont{Geim}},
  \bibinfo{journal}{Rev. Mod. Phys.} \textbf{\bibinfo{volume}{81}},
  \bibinfo{pages}{109} (\bibinfo{year}{2009}),
  \urlprefix\url{https://link.aps.org/doi/10.1103/RevModPhys.81.109}.

\bibitem[{\citenamefont{Pereira et~al.}(2009)\citenamefont{Pereira,
  Castro~Neto, and Peres}}]{PhysRevB.80.045401}
\bibinfo{author}{\bibfnamefont{V.~M.} \bibnamefont{Pereira}},
  \bibinfo{author}{\bibfnamefont{A.~H.} \bibnamefont{Castro~Neto}},
  \bibnamefont{and} \bibinfo{author}{\bibfnamefont{N.~M.~R.}
  \bibnamefont{Peres}}, \bibinfo{journal}{Phys. Rev. B}
  \textbf{\bibinfo{volume}{80}}, \bibinfo{pages}{045401}
  (\bibinfo{year}{2009}),
  \urlprefix\url{https://link.aps.org/doi/10.1103/PhysRevB.80.045401}.

\bibitem[{\citenamefont{Mohiuddin et~al.}(2009)\citenamefont{Mohiuddin,
  Lombardo, Nair, Bonetti, Savini, Jalil, Bonini, Basko, Galiotis, Marzari
  et~al.}}]{PhysRevB.79.205433}
\bibinfo{author}{\bibfnamefont{T.~M.~G.} \bibnamefont{Mohiuddin}},
  \bibinfo{author}{\bibfnamefont{A.}~\bibnamefont{Lombardo}},
  \bibinfo{author}{\bibfnamefont{R.~R.} \bibnamefont{Nair}},
  \bibinfo{author}{\bibfnamefont{A.}~\bibnamefont{Bonetti}},
  \bibinfo{author}{\bibfnamefont{G.}~\bibnamefont{Savini}},
  \bibinfo{author}{\bibfnamefont{R.}~\bibnamefont{Jalil}},
  \bibinfo{author}{\bibfnamefont{N.}~\bibnamefont{Bonini}},
  \bibinfo{author}{\bibfnamefont{D.~M.} \bibnamefont{Basko}},
  \bibinfo{author}{\bibfnamefont{C.}~\bibnamefont{Galiotis}},
  \bibinfo{author}{\bibfnamefont{N.}~\bibnamefont{Marzari}},
  \bibnamefont{et~al.}, \bibinfo{journal}{Phys. Rev. B}
  \textbf{\bibinfo{volume}{79}}, \bibinfo{pages}{205433}
  (\bibinfo{year}{2009}),
  \urlprefix\url{https://link.aps.org/doi/10.1103/PhysRevB.79.205433}.

\bibitem[{\citenamefont{Neek-Amal et~al.}(2013)\citenamefont{Neek-Amal, Covaci,
  Shakouri, and Peeters}}]{PhysRevB.88.115428}
\bibinfo{author}{\bibfnamefont{M.}~\bibnamefont{Neek-Amal}},
  \bibinfo{author}{\bibfnamefont{L.}~\bibnamefont{Covaci}},
  \bibinfo{author}{\bibfnamefont{K.}~\bibnamefont{Shakouri}}, \bibnamefont{and}
  \bibinfo{author}{\bibfnamefont{F.~M.} \bibnamefont{Peeters}},
  \bibinfo{journal}{Phys. Rev. B} \textbf{\bibinfo{volume}{88}},
  \bibinfo{pages}{115428} (\bibinfo{year}{2013}),
  \urlprefix\url{https://link.aps.org/doi/10.1103/PhysRevB.88.115428}.

\bibitem[{\citenamefont{Settnes et~al.}(2016)\citenamefont{Settnes, Power, and
  Jauho}}]{PhysRevB.93.035456}
\bibinfo{author}{\bibfnamefont{M.}~\bibnamefont{Settnes}},
  \bibinfo{author}{\bibfnamefont{S.~R.} \bibnamefont{Power}}, \bibnamefont{and}
  \bibinfo{author}{\bibfnamefont{A.-P.} \bibnamefont{Jauho}},
  \bibinfo{journal}{Phys. Rev. B} \textbf{\bibinfo{volume}{93}},
  \bibinfo{pages}{035456} (\bibinfo{year}{2016}),
  \urlprefix\url{https://link.aps.org/doi/10.1103/PhysRevB.93.035456}.

\bibitem[{\citenamefont{Guinea et~al.}(2010{\natexlab{b}})\citenamefont{Guinea,
  Geim, Katsnelson, and Novoselov}}]{PhysRevB.81.035408}
\bibinfo{author}{\bibfnamefont{F.}~\bibnamefont{Guinea}},
  \bibinfo{author}{\bibfnamefont{A.~K.} \bibnamefont{Geim}},
  \bibinfo{author}{\bibfnamefont{M.~I.} \bibnamefont{Katsnelson}},
  \bibnamefont{and} \bibinfo{author}{\bibfnamefont{K.~S.}
  \bibnamefont{Novoselov}}, \bibinfo{journal}{Phys. Rev. B}
  \textbf{\bibinfo{volume}{81}}, \bibinfo{pages}{035408}
  (\bibinfo{year}{2010}{\natexlab{b}}),
  \urlprefix\url{https://link.aps.org/doi/10.1103/PhysRevB.81.035408}.

\bibitem[{\citenamefont{Chang et~al.}(2012)\citenamefont{Chang, Albash, and
  Haas}}]{PhysRevB.86.125402}
\bibinfo{author}{\bibfnamefont{Y.}~\bibnamefont{Chang}},
  \bibinfo{author}{\bibfnamefont{T.}~\bibnamefont{Albash}}, \bibnamefont{and}
  \bibinfo{author}{\bibfnamefont{S.}~\bibnamefont{Haas}},
  \bibinfo{journal}{Phys. Rev. B} \textbf{\bibinfo{volume}{86}},
  \bibinfo{pages}{125402} (\bibinfo{year}{2012}),
  \urlprefix\url{https://link.aps.org/doi/10.1103/PhysRevB.86.125402}.

\bibitem[{\citenamefont{Zhang et~al.}(2014)\citenamefont{Zhang, Seifert, and
  Chang}}]{PhysRevLett.112.096805}
\bibinfo{author}{\bibfnamefont{D.-B.} \bibnamefont{Zhang}},
  \bibinfo{author}{\bibfnamefont{G.}~\bibnamefont{Seifert}}, \bibnamefont{and}
  \bibinfo{author}{\bibfnamefont{K.}~\bibnamefont{Chang}},
  \bibinfo{journal}{Phys. Rev. Lett.} \textbf{\bibinfo{volume}{112}},
  \bibinfo{pages}{096805} (\bibinfo{year}{2014}),
  \urlprefix\url{https://link.aps.org/doi/10.1103/PhysRevLett.112.096805}.

\bibitem[{\citenamefont{Ho et~al.}(2017)\citenamefont{Ho, Castro, and
  Cazalilla}}]{PhysRevB.96.155446}
\bibinfo{author}{\bibfnamefont{Y.-H.} \bibnamefont{Ho}},
  \bibinfo{author}{\bibfnamefont{E.~V.} \bibnamefont{Castro}},
  \bibnamefont{and} \bibinfo{author}{\bibfnamefont{M.~A.}
  \bibnamefont{Cazalilla}}, \bibinfo{journal}{Phys. Rev. B}
  \textbf{\bibinfo{volume}{96}}, \bibinfo{pages}{155446}
  (\bibinfo{year}{2017}),
  \urlprefix\url{https://link.aps.org/doi/10.1103/PhysRevB.96.155446}.

\bibitem[{\citenamefont{Levy et~al.}(2010)\citenamefont{Levy, Burke, Meaker,
  Panlasigui, Zettl, Guinea, Neto, and Crommie}}]{2010natureStrain}
\bibinfo{author}{\bibfnamefont{N.}~\bibnamefont{Levy}},
  \bibinfo{author}{\bibfnamefont{S.~A.} \bibnamefont{Burke}},
  \bibinfo{author}{\bibfnamefont{K.~L.} \bibnamefont{Meaker}},
  \bibinfo{author}{\bibfnamefont{M.}~\bibnamefont{Panlasigui}},
  \bibinfo{author}{\bibfnamefont{A.}~\bibnamefont{Zettl}},
  \bibinfo{author}{\bibfnamefont{F.}~\bibnamefont{Guinea}},
  \bibinfo{author}{\bibfnamefont{A.}~\bibnamefont{Neto}}, \bibnamefont{and}
  \bibinfo{author}{\bibfnamefont{M.~F.} \bibnamefont{Crommie}},
  \bibinfo{journal}{Science} \textbf{\bibinfo{volume}{329}},
  \bibinfo{pages}{P.544} (\bibinfo{year}{2010}).

\bibitem[{\citenamefont{Mao et~al.}(2020)\citenamefont{Mao, Milovanovi\'{c},
  Andelkovi\'{c}, Lai, Cao, Watanabe, Taniguchi, Covaci, Peeters, Geim
  et~al.}}]{RN45}
\bibinfo{author}{\bibfnamefont{J.}~\bibnamefont{Mao}},
  \bibinfo{author}{\bibfnamefont{S.~P.} \bibnamefont{Milovanovi\'{c}}},
  \bibinfo{author}{\bibfnamefont{M.}~\bibnamefont{Andelkovi\'{c}}},
  \bibinfo{author}{\bibfnamefont{X.}~\bibnamefont{Lai}},
  \bibinfo{author}{\bibfnamefont{Y.}~\bibnamefont{Cao}},
  \bibinfo{author}{\bibfnamefont{K.}~\bibnamefont{Watanabe}},
  \bibinfo{author}{\bibfnamefont{T.}~\bibnamefont{Taniguchi}},
  \bibinfo{author}{\bibfnamefont{L.}~\bibnamefont{Covaci}},
  \bibinfo{author}{\bibfnamefont{F.~M.} \bibnamefont{Peeters}},
  \bibinfo{author}{\bibfnamefont{A.~K.} \bibnamefont{Geim}},
  \bibnamefont{et~al.}, \bibinfo{journal}{Nature}
  \textbf{\bibinfo{volume}{584}}, \bibinfo{pages}{215} (\bibinfo{year}{2020}),
  ISSN \bibinfo{issn}{1476-4687},
  \urlprefix\url{https://doi.org/10.1038/s41586-020-2567-3}.

\bibitem[{\citenamefont{Meng et~al.}(2013)\citenamefont{Meng, He, Zheng, Liu,
  Yan, Yan, Chu, Bai, Dou, Zhang et~al.}}]{PhysRevB.87.205405}
\bibinfo{author}{\bibfnamefont{L.}~\bibnamefont{Meng}},
  \bibinfo{author}{\bibfnamefont{W.-Y.} \bibnamefont{He}},
  \bibinfo{author}{\bibfnamefont{H.}~\bibnamefont{Zheng}},
  \bibinfo{author}{\bibfnamefont{M.}~\bibnamefont{Liu}},
  \bibinfo{author}{\bibfnamefont{H.}~\bibnamefont{Yan}},
  \bibinfo{author}{\bibfnamefont{W.}~\bibnamefont{Yan}},
  \bibinfo{author}{\bibfnamefont{Z.-D.} \bibnamefont{Chu}},
  \bibinfo{author}{\bibfnamefont{K.}~\bibnamefont{Bai}},
  \bibinfo{author}{\bibfnamefont{R.-F.} \bibnamefont{Dou}},
  \bibinfo{author}{\bibfnamefont{Y.}~\bibnamefont{Zhang}},
  \bibnamefont{et~al.}, \bibinfo{journal}{Phys. Rev. B}
  \textbf{\bibinfo{volume}{87}}, \bibinfo{pages}{205405}
  (\bibinfo{year}{2013}),
  \urlprefix\url{https://link.aps.org/doi/10.1103/PhysRevB.87.205405}.

\bibitem[{\citenamefont{Pikulin et~al.}(2016)\citenamefont{Pikulin, Chen, and
  Franz}}]{PhysRevX.6.041021}
\bibinfo{author}{\bibfnamefont{D.~I.} \bibnamefont{Pikulin}},
  \bibinfo{author}{\bibfnamefont{A.}~\bibnamefont{Chen}}, \bibnamefont{and}
  \bibinfo{author}{\bibfnamefont{M.}~\bibnamefont{Franz}},
  \bibinfo{journal}{Phys. Rev. X} \textbf{\bibinfo{volume}{6}},
  \bibinfo{pages}{041021} (\bibinfo{year}{2016}),
  \urlprefix\url{https://link.aps.org/doi/10.1103/PhysRevX.6.041021}.

\bibitem[{\citenamefont{Liu et~al.}(2017{\natexlab{a}})\citenamefont{Liu,
  Franz, and Fujimoto}}]{PhysRevB.96.224518}
\bibinfo{author}{\bibfnamefont{T.}~\bibnamefont{Liu}},
  \bibinfo{author}{\bibfnamefont{M.}~\bibnamefont{Franz}}, \bibnamefont{and}
  \bibinfo{author}{\bibfnamefont{S.}~\bibnamefont{Fujimoto}},
  \bibinfo{journal}{Phys. Rev. B} \textbf{\bibinfo{volume}{96}},
  \bibinfo{pages}{224518} (\bibinfo{year}{2017}{\natexlab{a}}),
  \urlprefix\url{https://link.aps.org/doi/10.1103/PhysRevB.96.224518}.

\bibitem[{\citenamefont{Liu et~al.}(2017{\natexlab{b}})\citenamefont{Liu,
  Pikulin, and Franz}}]{PhysRevB.95.041201}
\bibinfo{author}{\bibfnamefont{T.}~\bibnamefont{Liu}},
  \bibinfo{author}{\bibfnamefont{D.~I.} \bibnamefont{Pikulin}},
  \bibnamefont{and} \bibinfo{author}{\bibfnamefont{M.}~\bibnamefont{Franz}},
  \bibinfo{journal}{Phys. Rev. B} \textbf{\bibinfo{volume}{95}},
  \bibinfo{pages}{041201} (\bibinfo{year}{2017}{\natexlab{b}}),
  \urlprefix\url{https://link.aps.org/doi/10.1103/PhysRevB.95.041201}.

\bibitem[{\citenamefont{Nica and Franz}(2018)}]{PhysRevB.97.024520}
\bibinfo{author}{\bibfnamefont{E.~M.} \bibnamefont{Nica}} \bibnamefont{and}
  \bibinfo{author}{\bibfnamefont{M.}~\bibnamefont{Franz}},
  \bibinfo{journal}{Phys. Rev. B} \textbf{\bibinfo{volume}{97}},
  \bibinfo{pages}{024520} (\bibinfo{year}{2018}),
  \urlprefix\url{https://link.aps.org/doi/10.1103/PhysRevB.97.024520}.

\bibitem[{\citenamefont{Li and Kovalev}(2020)}]{PhysRevLett.125.257201}
\bibinfo{author}{\bibfnamefont{B.}~\bibnamefont{Li}} \bibnamefont{and}
  \bibinfo{author}{\bibfnamefont{A.~A.} \bibnamefont{Kovalev}},
  \bibinfo{journal}{Phys. Rev. Lett.} \textbf{\bibinfo{volume}{125}},
  \bibinfo{pages}{257201} (\bibinfo{year}{2020}),
  \urlprefix\url{https://link.aps.org/doi/10.1103/PhysRevLett.125.257201}.

\bibitem[{\citenamefont{Liu and Shi}(2021)}]{PhysRevB.103.144420}
\bibinfo{author}{\bibfnamefont{T.}~\bibnamefont{Liu}} \bibnamefont{and}
  \bibinfo{author}{\bibfnamefont{Z.}~\bibnamefont{Shi}},
  \bibinfo{journal}{Phys. Rev. B} \textbf{\bibinfo{volume}{103}},
  \bibinfo{pages}{144420} (\bibinfo{year}{2021}),
  \urlprefix\url{https://link.aps.org/doi/10.1103/PhysRevB.103.144420}.

\bibitem[{\citenamefont{Rachel et~al.}(2016)\citenamefont{Rachel, Fritz, and
  Vojta}}]{PhysRevLett.116.167201}
\bibinfo{author}{\bibfnamefont{S.}~\bibnamefont{Rachel}},
  \bibinfo{author}{\bibfnamefont{L.}~\bibnamefont{Fritz}}, \bibnamefont{and}
  \bibinfo{author}{\bibfnamefont{M.}~\bibnamefont{Vojta}},
  \bibinfo{journal}{Phys. Rev. Lett.} \textbf{\bibinfo{volume}{116}},
  \bibinfo{pages}{167201} (\bibinfo{year}{2016}),
  \urlprefix\url{https://link.aps.org/doi/10.1103/PhysRevLett.116.167201}.

\bibitem[{\citenamefont{Guglielmon et~al.}(2021)\citenamefont{Guglielmon,
  Rechtsman, and Weinstein}}]{PhysRevA.103.013505}
\bibinfo{author}{\bibfnamefont{J.}~\bibnamefont{Guglielmon}},
  \bibinfo{author}{\bibfnamefont{M.~C.} \bibnamefont{Rechtsman}},
  \bibnamefont{and} \bibinfo{author}{\bibfnamefont{M.~I.}
  \bibnamefont{Weinstein}}, \bibinfo{journal}{Phys. Rev. A}
  \textbf{\bibinfo{volume}{103}}, \bibinfo{pages}{013505}
  (\bibinfo{year}{2021}),
  \urlprefix\url{https://link.aps.org/doi/10.1103/PhysRevA.103.013505}.

\bibitem[{\citenamefont{Yang et~al.}(2017)\citenamefont{Yang, Gao, Yang, and
  Zhang}}]{PhysRevLett.118.194301}
\bibinfo{author}{\bibfnamefont{Z.}~\bibnamefont{Yang}},
  \bibinfo{author}{\bibfnamefont{F.}~\bibnamefont{Gao}},
  \bibinfo{author}{\bibfnamefont{Y.}~\bibnamefont{Yang}}, \bibnamefont{and}
  \bibinfo{author}{\bibfnamefont{B.}~\bibnamefont{Zhang}},
  \bibinfo{journal}{Phys. Rev. Lett.} \textbf{\bibinfo{volume}{118}},
  \bibinfo{pages}{194301} (\bibinfo{year}{2017}),
  \urlprefix\url{https://link.aps.org/doi/10.1103/PhysRevLett.118.194301}.

\bibitem[{\citenamefont{Nayga et~al.}(2019)\citenamefont{Nayga, Rachel, and
  Vojta}}]{PhysRevLett.123.207204}
\bibinfo{author}{\bibfnamefont{M.~M.} \bibnamefont{Nayga}},
  \bibinfo{author}{\bibfnamefont{S.}~\bibnamefont{Rachel}}, \bibnamefont{and}
  \bibinfo{author}{\bibfnamefont{M.}~\bibnamefont{Vojta}},
  \bibinfo{journal}{Phys. Rev. Lett.} \textbf{\bibinfo{volume}{123}},
  \bibinfo{pages}{207204} (\bibinfo{year}{2019}),
  \urlprefix\url{https://link.aps.org/doi/10.1103/PhysRevLett.123.207204}.

\bibitem[{\citenamefont{Sun et~al.}(2021)\citenamefont{Sun, Ma, Ying, Guo, and
  Feng}}]{sun2021quantum}
\bibinfo{author}{\bibfnamefont{J.}~\bibnamefont{Sun}},
  \bibinfo{author}{\bibfnamefont{N.}~\bibnamefont{Ma}},
  \bibinfo{author}{\bibfnamefont{T.}~\bibnamefont{Ying}},
  \bibinfo{author}{\bibfnamefont{H.}~\bibnamefont{Guo}}, \bibnamefont{and}
  \bibinfo{author}{\bibfnamefont{S.}~\bibnamefont{Feng}},
  \emph{\bibinfo{title}{Quantum monte carlo study of honeycomb antiferromagnets
  under a triaxial strain}} (\bibinfo{year}{2021}), \eprint{2106.04358},
  \urlprefix\url{https://arxiv.org/abs/2106.04358}.

\bibitem[{\citenamefont{Lantagne-Hurtubise
  et~al.}(2020)\citenamefont{Lantagne-Hurtubise, Zhang, and
  Franz}}]{PhysRevB.101.085423}
\bibinfo{author}{\bibfnamefont{E.}~\bibnamefont{Lantagne-Hurtubise}},
  \bibinfo{author}{\bibfnamefont{X.-X.} \bibnamefont{Zhang}}, \bibnamefont{and}
  \bibinfo{author}{\bibfnamefont{M.}~\bibnamefont{Franz}},
  \bibinfo{journal}{Phys. Rev. B} \textbf{\bibinfo{volume}{101}},
  \bibinfo{pages}{085423} (\bibinfo{year}{2020}),
  \urlprefix\url{https://link.aps.org/doi/10.1103/PhysRevB.101.085423}.

\bibitem[{\citenamefont{Sylju\aa{}sen and Sandvik}(2002)}]{sandvik2002}
\bibinfo{author}{\bibfnamefont{O.~F.} \bibnamefont{Sylju\aa{}sen}}
  \bibnamefont{and} \bibinfo{author}{\bibfnamefont{A.~W.}
  \bibnamefont{Sandvik}}, \bibinfo{journal}{Phys. Rev. E}
  \textbf{\bibinfo{volume}{66}}, \bibinfo{pages}{046701}
  (\bibinfo{year}{2002}),
  \urlprefix\url{https://link.aps.org/doi/10.1103/PhysRevE.66.046701}.

\bibitem[{\citenamefont{Sylju\aa{}sen}(2003)}]{syljuasen2003}
\bibinfo{author}{\bibfnamefont{O.~F.} \bibnamefont{Sylju\aa{}sen}},
  \bibinfo{journal}{Phys. Rev. E} \textbf{\bibinfo{volume}{67}},
  \bibinfo{pages}{046701} (\bibinfo{year}{2003}),
  \urlprefix\url{https://link.aps.org/doi/10.1103/PhysRevE.67.046701}.

\bibitem[{\citenamefont{Bauer et~al.}(2011)\citenamefont{Bauer, Carr, Evertz,
  Feiguin, Freire, Fuchs, Gamper, Gukelberger, Gull, Guertler
  et~al.}}]{Bauer2011}
\bibinfo{author}{\bibfnamefont{B.}~\bibnamefont{Bauer}},
  \bibinfo{author}{\bibfnamefont{L.~D.} \bibnamefont{Carr}},
  \bibinfo{author}{\bibfnamefont{H.~G.} \bibnamefont{Evertz}},
  \bibinfo{author}{\bibfnamefont{A.}~\bibnamefont{Feiguin}},
  \bibinfo{author}{\bibfnamefont{J.}~\bibnamefont{Freire}},
  \bibinfo{author}{\bibfnamefont{S.}~\bibnamefont{Fuchs}},
  \bibinfo{author}{\bibfnamefont{L.}~\bibnamefont{Gamper}},
  \bibinfo{author}{\bibfnamefont{J.}~\bibnamefont{Gukelberger}},
  \bibinfo{author}{\bibfnamefont{E.}~\bibnamefont{Gull}},
  \bibinfo{author}{\bibfnamefont{S.}~\bibnamefont{Guertler}},
  \bibnamefont{et~al.}, \bibinfo{journal}{Journal of Statistical Mechanics:
  Theory and Experiment} \textbf{\bibinfo{volume}{2011}},
  \bibinfo{pages}{P05001} (\bibinfo{year}{2011}),
  \urlprefix\url{https://doi.org/10.1088%2F1742-5468%2F2011%2F05%2Fp05001}.

\bibitem[{\citenamefont{Alet et~al.}(2005)\citenamefont{Alet, Wessel, and
  Troyer}}]{fabien2005}
\bibinfo{author}{\bibfnamefont{F.}~\bibnamefont{Alet}},
  \bibinfo{author}{\bibfnamefont{S.}~\bibnamefont{Wessel}}, \bibnamefont{and}
  \bibinfo{author}{\bibfnamefont{M.}~\bibnamefont{Troyer}},
  \bibinfo{journal}{Phys. Rev. E} \textbf{\bibinfo{volume}{71}},
  \bibinfo{pages}{036706} (\bibinfo{year}{2005}),
  \urlprefix\url{https://link.aps.org/doi/10.1103/PhysRevE.71.036706}.

\bibitem[{\citenamefont{Pollet et~al.}(2004)\citenamefont{Pollet, Rombouts,
  Van~Houcke, and Heyde}}]{pollet2004}
\bibinfo{author}{\bibfnamefont{L.}~\bibnamefont{Pollet}},
  \bibinfo{author}{\bibfnamefont{S.~M.~A.} \bibnamefont{Rombouts}},
  \bibinfo{author}{\bibfnamefont{K.}~\bibnamefont{Van~Houcke}},
  \bibnamefont{and} \bibinfo{author}{\bibfnamefont{K.}~\bibnamefont{Heyde}},
  \bibinfo{journal}{Phys. Rev. E} \textbf{\bibinfo{volume}{70}},
  \bibinfo{pages}{056705} (\bibinfo{year}{2004}),
  \urlprefix\url{https://link.aps.org/doi/10.1103/PhysRevE.70.056705}.

\bibitem[{\citenamefont{Holstein and Primakoff}(1940)}]{hptransformation}
\bibinfo{author}{\bibfnamefont{T.}~\bibnamefont{Holstein}} \bibnamefont{and}
  \bibinfo{author}{\bibfnamefont{H.}~\bibnamefont{Primakoff}},
  \bibinfo{journal}{Phys. Rev.} \textbf{\bibinfo{volume}{58}},
  \bibinfo{pages}{1098} (\bibinfo{year}{1940}),
  \urlprefix\url{https://link.aps.org/doi/10.1103/PhysRev.58.1098}.

\bibitem[{\citenamefont{White et~al.}(1965)\citenamefont{White, Sparks, and
  Ortenburger}}]{PhysRev.139.A450}
\bibinfo{author}{\bibfnamefont{R.~M.} \bibnamefont{White}},
  \bibinfo{author}{\bibfnamefont{M.}~\bibnamefont{Sparks}}, \bibnamefont{and}
  \bibinfo{author}{\bibfnamefont{I.}~\bibnamefont{Ortenburger}},
  \bibinfo{journal}{Phys. Rev.} \textbf{\bibinfo{volume}{139}},
  \bibinfo{pages}{A450} (\bibinfo{year}{1965}),
  \urlprefix\url{https://link.aps.org/doi/10.1103/PhysRev.139.A450}.

\bibitem[{\citenamefont{Xiao}(2009)}]{xiao2009theory}
\bibinfo{author}{\bibfnamefont{M.-W.} \bibnamefont{Xiao}},
  \bibinfo{journal}{arXiv:0908.0787}  (\bibinfo{year}{2009}).

\bibitem[{\citenamefont{Huang et~al.}(2017)\citenamefont{Huang, Hikihara, Lee,
  and Lin}}]{2017Edge}
\bibinfo{author}{\bibfnamefont{W.~M.} \bibnamefont{Huang}},
  \bibinfo{author}{\bibfnamefont{T.}~\bibnamefont{Hikihara}},
  \bibinfo{author}{\bibfnamefont{Y.~C.} \bibnamefont{Lee}}, \bibnamefont{and}
  \bibinfo{author}{\bibfnamefont{H.~H.} \bibnamefont{Lin}},
  \bibinfo{journal}{Scientific Reports} \textbf{\bibinfo{volume}{7}},
  \bibinfo{pages}{43678} (\bibinfo{year}{2017}).

\bibitem[{\citenamefont{Sala et~al.}(2021)\citenamefont{Sala, Stone, Rai, May,
  Laurell, Garlea, Butch, Lumsden, Ehlers, and Pokharel}}]{VanHsingularity}
\bibinfo{author}{\bibfnamefont{G.}~\bibnamefont{Sala}},
  \bibinfo{author}{\bibfnamefont{M.~B.} \bibnamefont{Stone}},
  \bibinfo{author}{\bibfnamefont{B.~K.} \bibnamefont{Rai}},
  \bibinfo{author}{\bibfnamefont{A.~F.} \bibnamefont{May}},
  \bibinfo{author}{\bibfnamefont{P.}~\bibnamefont{Laurell}},
  \bibinfo{author}{\bibfnamefont{V.~O.} \bibnamefont{Garlea}},
  \bibinfo{author}{\bibfnamefont{N.~P.} \bibnamefont{Butch}},
  \bibinfo{author}{\bibfnamefont{M.}~\bibnamefont{Lumsden}},
  \bibinfo{author}{\bibfnamefont{G.}~\bibnamefont{Ehlers}}, \bibnamefont{and}
  \bibinfo{author}{\bibfnamefont{G.}~\bibnamefont{Pokharel}},
  \bibinfo{journal}{Nature Communications} \textbf{\bibinfo{volume}{12}},
  \bibinfo{pages}{171} (\bibinfo{year}{2021}),
  \urlprefix\url{https://doi.org/10.1038/s41467-020-20335-5}.

\bibitem[{\citenamefont{Goerbig}(2011)}]{RevModPhys.83.1193}
\bibinfo{author}{\bibfnamefont{M.~O.} \bibnamefont{Goerbig}},
  \bibinfo{journal}{Rev. Mod. Phys.} \textbf{\bibinfo{volume}{83}},
  \bibinfo{pages}{1193} (\bibinfo{year}{2011}),
  \urlprefix\url{https://link.aps.org/doi/10.1103/RevModPhys.83.1193}.

\bibitem[{\citenamefont{Wessel and Milat}(2005)}]{PhysRevB.71.104427}
\bibinfo{author}{\bibfnamefont{S.}~\bibnamefont{Wessel}} \bibnamefont{and}
  \bibinfo{author}{\bibfnamefont{I.}~\bibnamefont{Milat}},
  \bibinfo{journal}{Phys. Rev. B} \textbf{\bibinfo{volume}{71}},
  \bibinfo{pages}{104427} (\bibinfo{year}{2005}),
  \urlprefix\url{https://link.aps.org/doi/10.1103/PhysRevB.71.104427}.

\bibitem[{\citenamefont{Hikihara et~al.}(2003)\citenamefont{Hikihara, Hu, Lin,
  and Mou}}]{PhysRevB.68.035432}
\bibinfo{author}{\bibfnamefont{T.}~\bibnamefont{Hikihara}},
  \bibinfo{author}{\bibfnamefont{X.}~\bibnamefont{Hu}},
  \bibinfo{author}{\bibfnamefont{H.-H.} \bibnamefont{Lin}}, \bibnamefont{and}
  \bibinfo{author}{\bibfnamefont{C.-Y.} \bibnamefont{Mou}},
  \bibinfo{journal}{Phys. Rev. B} \textbf{\bibinfo{volume}{68}},
  \bibinfo{pages}{035432} (\bibinfo{year}{2003}),
  \urlprefix\url{https://link.aps.org/doi/10.1103/PhysRevB.68.035432}.

\bibitem[{\citenamefont{Golor et~al.}(2014)\citenamefont{Golor, Wessel, and
  Schmidt}}]{PhysRevLett.112.046601}
\bibinfo{author}{\bibfnamefont{M.}~\bibnamefont{Golor}},
  \bibinfo{author}{\bibfnamefont{S.}~\bibnamefont{Wessel}}, \bibnamefont{and}
  \bibinfo{author}{\bibfnamefont{M.~J.} \bibnamefont{Schmidt}},
  \bibinfo{journal}{Phys. Rev. Lett.} \textbf{\bibinfo{volume}{112}},
  \bibinfo{pages}{046601} (\bibinfo{year}{2014}),
  \urlprefix\url{https://link.aps.org/doi/10.1103/PhysRevLett.112.046601}.

\bibitem[{\citenamefont{Golor et~al.}(2013)\citenamefont{Golor, Lang, and
  Wessel}}]{PhysRevB.87.155441}
\bibinfo{author}{\bibfnamefont{M.}~\bibnamefont{Golor}},
  \bibinfo{author}{\bibfnamefont{T.~C.} \bibnamefont{Lang}}, \bibnamefont{and}
  \bibinfo{author}{\bibfnamefont{S.}~\bibnamefont{Wessel}},
  \bibinfo{journal}{Phys. Rev. B} \textbf{\bibinfo{volume}{87}},
  \bibinfo{pages}{155441} (\bibinfo{year}{2013}),
  \urlprefix\url{https://link.aps.org/doi/10.1103/PhysRevB.87.155441}.

\bibitem[{\citenamefont{Feldner et~al.}(2010)\citenamefont{Feldner, Meng,
  Honecker, Cabra, Wessel, and Assaad}}]{PhysRevB.81.115416}
\bibinfo{author}{\bibfnamefont{H.}~\bibnamefont{Feldner}},
  \bibinfo{author}{\bibfnamefont{Z.~Y.} \bibnamefont{Meng}},
  \bibinfo{author}{\bibfnamefont{A.}~\bibnamefont{Honecker}},
  \bibinfo{author}{\bibfnamefont{D.}~\bibnamefont{Cabra}},
  \bibinfo{author}{\bibfnamefont{S.}~\bibnamefont{Wessel}}, \bibnamefont{and}
  \bibinfo{author}{\bibfnamefont{F.~F.} \bibnamefont{Assaad}},
  \bibinfo{journal}{Phys. Rev. B} \textbf{\bibinfo{volume}{81}},
  \bibinfo{pages}{115416} (\bibinfo{year}{2010}),
  \urlprefix\url{https://link.aps.org/doi/10.1103/PhysRevB.81.115416}.

\bibitem[{\citenamefont{Roy et~al.}(2014)\citenamefont{Roy, Assaad, and
  Herbut}}]{PhysRevX.4.021042}
\bibinfo{author}{\bibfnamefont{B.}~\bibnamefont{Roy}},
  \bibinfo{author}{\bibfnamefont{F.~F.} \bibnamefont{Assaad}},
  \bibnamefont{and} \bibinfo{author}{\bibfnamefont{I.~F.}
  \bibnamefont{Herbut}}, \bibinfo{journal}{Phys. Rev. X}
  \textbf{\bibinfo{volume}{4}}, \bibinfo{pages}{021042} (\bibinfo{year}{2014}),
  \urlprefix\url{https://link.aps.org/doi/10.1103/PhysRevX.4.021042}.

\bibitem[{\citenamefont{Wessel et~al.}(2003)\citenamefont{Wessel, Jagannathan,
  and Haas}}]{PhysRevLett.90.177205}
\bibinfo{author}{\bibfnamefont{S.}~\bibnamefont{Wessel}},
  \bibinfo{author}{\bibfnamefont{A.}~\bibnamefont{Jagannathan}},
  \bibnamefont{and} \bibinfo{author}{\bibfnamefont{S.}~\bibnamefont{Haas}},
  \bibinfo{journal}{Phys. Rev. Lett.} \textbf{\bibinfo{volume}{90}},
  \bibinfo{pages}{177205} (\bibinfo{year}{2003}),
  \urlprefix\url{https://link.aps.org/doi/10.1103/PhysRevLett.90.177205}.

\bibitem[{\citenamefont{Dalla~Piazza et~al.}(2015)\citenamefont{Dalla~Piazza,
  Mourigal, Christensen, Nilsen, Tregenna-Piggott, Perring, Enderle, McMorrow,
  Ivanov, and R{\o}nnow}}]{dalla2015fractional}
\bibinfo{author}{\bibfnamefont{B.}~\bibnamefont{Dalla~Piazza}},
  \bibinfo{author}{\bibfnamefont{M.}~\bibnamefont{Mourigal}},
  \bibinfo{author}{\bibfnamefont{N.~B.} \bibnamefont{Christensen}},
  \bibinfo{author}{\bibfnamefont{G.}~\bibnamefont{Nilsen}},
  \bibinfo{author}{\bibfnamefont{P.}~\bibnamefont{Tregenna-Piggott}},
  \bibinfo{author}{\bibfnamefont{T.}~\bibnamefont{Perring}},
  \bibinfo{author}{\bibfnamefont{M.}~\bibnamefont{Enderle}},
  \bibinfo{author}{\bibfnamefont{D.~F.} \bibnamefont{McMorrow}},
  \bibinfo{author}{\bibfnamefont{D.}~\bibnamefont{Ivanov}}, \bibnamefont{and}
  \bibinfo{author}{\bibfnamefont{H.~M.} \bibnamefont{R{\o}nnow}},
  \bibinfo{journal}{Nature physics} \textbf{\bibinfo{volume}{11}},
  \bibinfo{pages}{62} (\bibinfo{year}{2015}).

\bibitem[{\citenamefont{Shao et~al.}(2017)\citenamefont{Shao, Qin, Capponi,
  Chesi, Meng, and Sandvik}}]{PhysRevX.7.041072}
\bibinfo{author}{\bibfnamefont{H.}~\bibnamefont{Shao}},
  \bibinfo{author}{\bibfnamefont{Y.~Q.} \bibnamefont{Qin}},
  \bibinfo{author}{\bibfnamefont{S.}~\bibnamefont{Capponi}},
  \bibinfo{author}{\bibfnamefont{S.}~\bibnamefont{Chesi}},
  \bibinfo{author}{\bibfnamefont{Z.~Y.} \bibnamefont{Meng}}, \bibnamefont{and}
  \bibinfo{author}{\bibfnamefont{A.~W.} \bibnamefont{Sandvik}},
  \bibinfo{journal}{Phys. Rev. X} \textbf{\bibinfo{volume}{7}},
  \bibinfo{pages}{041072} (\bibinfo{year}{2017}),
  \urlprefix\url{https://link.aps.org/doi/10.1103/PhysRevX.7.041072}.

\bibitem[{\citenamefont{Gomez-Santos and
  Joannopoulos}(1987)}]{Joannopoulos1987}
\bibinfo{author}{\bibfnamefont{G.}~\bibnamefont{Gomez-Santos}}
  \bibnamefont{and} \bibinfo{author}{\bibfnamefont{J.~D.}
  \bibnamefont{Joannopoulos}}, \bibinfo{journal}{Phys. Rev. B}
  \textbf{\bibinfo{volume}{36}}, \bibinfo{pages}{8707} (\bibinfo{year}{1987}),
  \urlprefix\url{https://link.aps.org/doi/10.1103/PhysRevB.36.8707}.

\bibitem[{\citenamefont{Guo et~al.}(2021)\citenamefont{Guo, Sun, Zhu, Li, Guo,
  and Feng}}]{guo2021quantum}
\bibinfo{author}{\bibfnamefont{J.}~\bibnamefont{Guo}},
  \bibinfo{author}{\bibfnamefont{J.}~\bibnamefont{Sun}},
  \bibinfo{author}{\bibfnamefont{X.}~\bibnamefont{Zhu}},
  \bibinfo{author}{\bibfnamefont{C.-A.} \bibnamefont{Li}},
  \bibinfo{author}{\bibfnamefont{H.}~\bibnamefont{Guo}}, \bibnamefont{and}
  \bibinfo{author}{\bibfnamefont{S.}~\bibnamefont{Feng}}
  (\bibinfo{year}{2021}), \eprint{2010.05402}.

\bibitem[{\citenamefont{Zhang et~al.}(2021)\citenamefont{Zhang, Zhang, Ni,
  Yang, Xiang, and Gong}}]{Yi-LinZhang:27501}
\bibinfo{author}{\bibfnamefont{Y.-L.} \bibnamefont{Zhang}},
  \bibinfo{author}{\bibfnamefont{Y.-Y.} \bibnamefont{Zhang}},
  \bibinfo{author}{\bibfnamefont{J.-Y.} \bibnamefont{Ni}},
  \bibinfo{author}{\bibfnamefont{J.-H.} \bibnamefont{Yang}},
  \bibinfo{author}{\bibfnamefont{H.-J.} \bibnamefont{Xiang}}, \bibnamefont{and}
  \bibinfo{author}{\bibfnamefont{X.-G.} \bibnamefont{Gong}},
  \bibinfo{journal}{Chinese Physics Letters} \textbf{\bibinfo{volume}{38}},
  \bibinfo{eid}{027501} (\bibinfo{year}{2021}),
  \urlprefix\url{http://cpl.iphy.ac.cn/EN/abstract/article_105845.shtml}.

\bibitem[{\citenamefont{Burch et~al.}(2018)\citenamefont{Burch, Mandrus, and
  Park}}]{burch2018magnetism}
\bibinfo{author}{\bibfnamefont{K.~S.} \bibnamefont{Burch}},
  \bibinfo{author}{\bibfnamefont{D.}~\bibnamefont{Mandrus}}, \bibnamefont{and}
  \bibinfo{author}{\bibfnamefont{J.-G.} \bibnamefont{Park}},
  \bibinfo{journal}{Nature} \textbf{\bibinfo{volume}{563}}, \bibinfo{pages}{47}
  (\bibinfo{year}{2018}).

\bibitem[{\citenamefont{Gibertini et~al.}(2019)\citenamefont{Gibertini,
  Koperski, Morpurgo, and Novoselov}}]{gibertini2019magnetic}
\bibinfo{author}{\bibfnamefont{M.}~\bibnamefont{Gibertini}},
  \bibinfo{author}{\bibfnamefont{M.}~\bibnamefont{Koperski}},
  \bibinfo{author}{\bibfnamefont{A.}~\bibnamefont{Morpurgo}}, \bibnamefont{and}
  \bibinfo{author}{\bibfnamefont{K.}~\bibnamefont{Novoselov}},
  \bibinfo{journal}{Nature nanotechnology} \textbf{\bibinfo{volume}{14}},
  \bibinfo{pages}{408} (\bibinfo{year}{2019}).

\bibitem[{\citenamefont{Yuan et~al.}(2020)\citenamefont{Yuan, Khait, Shu, Chou,
  Stone, Clancy, Paramekanti, and Kim}}]{PhysRevX.10.011062}
\bibinfo{author}{\bibfnamefont{B.}~\bibnamefont{Yuan}},
  \bibinfo{author}{\bibfnamefont{I.}~\bibnamefont{Khait}},
  \bibinfo{author}{\bibfnamefont{G.-J.} \bibnamefont{Shu}},
  \bibinfo{author}{\bibfnamefont{F.~C.} \bibnamefont{Chou}},
  \bibinfo{author}{\bibfnamefont{M.~B.} \bibnamefont{Stone}},
  \bibinfo{author}{\bibfnamefont{J.~P.} \bibnamefont{Clancy}},
  \bibinfo{author}{\bibfnamefont{A.}~\bibnamefont{Paramekanti}},
  \bibnamefont{and} \bibinfo{author}{\bibfnamefont{Y.-J.} \bibnamefont{Kim}},
  \bibinfo{journal}{Phys. Rev. X} \textbf{\bibinfo{volume}{10}},
  \bibinfo{pages}{011062} (\bibinfo{year}{2020}),
  \urlprefix\url{https://link.aps.org/doi/10.1103/PhysRevX.10.011062}.

\bibitem[{\citenamefont{Lee et~al.}(2016)\citenamefont{Lee, Lee, Ryoo, Kang,
  Kim, Kim, Park, Park, and Cheong}}]{material_1}
\bibinfo{author}{\bibfnamefont{J.-U.} \bibnamefont{Lee}},
  \bibinfo{author}{\bibfnamefont{S.}~\bibnamefont{Lee}},
  \bibinfo{author}{\bibfnamefont{J.~H.} \bibnamefont{Ryoo}},
  \bibinfo{author}{\bibfnamefont{S.}~\bibnamefont{Kang}},
  \bibinfo{author}{\bibfnamefont{T.~Y.} \bibnamefont{Kim}},
  \bibinfo{author}{\bibfnamefont{P.}~\bibnamefont{Kim}},
  \bibinfo{author}{\bibfnamefont{C.-H.} \bibnamefont{Park}},
  \bibinfo{author}{\bibfnamefont{J.-G.} \bibnamefont{Park}}, \bibnamefont{and}
  \bibinfo{author}{\bibfnamefont{H.}~\bibnamefont{Cheong}},
  \bibinfo{journal}{Nano Letters} \textbf{\bibinfo{volume}{16}},
  \bibinfo{pages}{7433} (\bibinfo{year}{2016}),
  \urlprefix\url{https://doi.org/10.1021/acs.nanolett.6b03052}.

\bibitem[{\citenamefont{Wang et~al.}(2016)\citenamefont{Wang, Du, Liu, Hu,
  Zhang, Zhang, Owen, Lu, Gan, Sengupta et~al.}}]{Wang_2016}
\bibinfo{author}{\bibfnamefont{X.}~\bibnamefont{Wang}},
  \bibinfo{author}{\bibfnamefont{K.}~\bibnamefont{Du}},
  \bibinfo{author}{\bibfnamefont{Y.~Y.~F.} \bibnamefont{Liu}},
  \bibinfo{author}{\bibfnamefont{P.}~\bibnamefont{Hu}},
  \bibinfo{author}{\bibfnamefont{J.}~\bibnamefont{Zhang}},
  \bibinfo{author}{\bibfnamefont{Q.}~\bibnamefont{Zhang}},
  \bibinfo{author}{\bibfnamefont{M.~H.~S.} \bibnamefont{Owen}},
  \bibinfo{author}{\bibfnamefont{X.}~\bibnamefont{Lu}},
  \bibinfo{author}{\bibfnamefont{C.~K.} \bibnamefont{Gan}},
  \bibinfo{author}{\bibfnamefont{P.}~\bibnamefont{Sengupta}},
  \bibnamefont{et~al.}, \bibinfo{journal}{2D Materials}
  \textbf{\bibinfo{volume}{3}}, \bibinfo{pages}{031009} (\bibinfo{year}{2016}),
  \urlprefix\url{https://doi.org/10.1088/2053-1583/3/3/031009}.

\bibitem[{\citenamefont{Gong et~al.}(2017)\citenamefont{Gong, Li, Li, Ji,
  Stern, Xia, Cao, Bao, Wang, Wang et~al.}}]{gong2017discovery}
\bibinfo{author}{\bibfnamefont{C.}~\bibnamefont{Gong}},
  \bibinfo{author}{\bibfnamefont{L.}~\bibnamefont{Li}},
  \bibinfo{author}{\bibfnamefont{Z.}~\bibnamefont{Li}},
  \bibinfo{author}{\bibfnamefont{H.}~\bibnamefont{Ji}},
  \bibinfo{author}{\bibfnamefont{A.}~\bibnamefont{Stern}},
  \bibinfo{author}{\bibfnamefont{Y.}~\bibnamefont{Xia}},
  \bibinfo{author}{\bibfnamefont{T.}~\bibnamefont{Cao}},
  \bibinfo{author}{\bibfnamefont{W.}~\bibnamefont{Bao}},
  \bibinfo{author}{\bibfnamefont{C.}~\bibnamefont{Wang}},
  \bibinfo{author}{\bibfnamefont{Y.}~\bibnamefont{Wang}}, \bibnamefont{et~al.},
  \bibinfo{journal}{Nature} \textbf{\bibinfo{volume}{546}},
  \bibinfo{pages}{265} (\bibinfo{year}{2017}).

\bibitem[{\citenamefont{Bonilla et~al.}(2018)\citenamefont{Bonilla, Kolekar,
  Ma, Diaz, Kalappattil, Das, Eggers, Gutierrez, Phan, and
  Batzill}}]{bonilla2018strong}
\bibinfo{author}{\bibfnamefont{M.}~\bibnamefont{Bonilla}},
  \bibinfo{author}{\bibfnamefont{S.}~\bibnamefont{Kolekar}},
  \bibinfo{author}{\bibfnamefont{Y.}~\bibnamefont{Ma}},
  \bibinfo{author}{\bibfnamefont{H.~C.} \bibnamefont{Diaz}},
  \bibinfo{author}{\bibfnamefont{V.}~\bibnamefont{Kalappattil}},
  \bibinfo{author}{\bibfnamefont{R.}~\bibnamefont{Das}},
  \bibinfo{author}{\bibfnamefont{T.}~\bibnamefont{Eggers}},
  \bibinfo{author}{\bibfnamefont{H.~R.} \bibnamefont{Gutierrez}},
  \bibinfo{author}{\bibfnamefont{M.-H.} \bibnamefont{Phan}}, \bibnamefont{and}
  \bibinfo{author}{\bibfnamefont{M.}~\bibnamefont{Batzill}},
  \bibinfo{journal}{Nature nanotechnology} \textbf{\bibinfo{volume}{13}},
  \bibinfo{pages}{289} (\bibinfo{year}{2018}).

\bibitem[{\citenamefont{O’Hara et~al.}(2018)\citenamefont{O’Hara, Zhu,
  Trout, Ahmed, Luo, Lee, Brenner, Rajan, Gupta, McComb et~al.}}]{material_2}
\bibinfo{author}{\bibfnamefont{D.~J.} \bibnamefont{O’Hara}},
  \bibinfo{author}{\bibfnamefont{T.}~\bibnamefont{Zhu}},
  \bibinfo{author}{\bibfnamefont{A.~H.} \bibnamefont{Trout}},
  \bibinfo{author}{\bibfnamefont{A.~S.} \bibnamefont{Ahmed}},
  \bibinfo{author}{\bibfnamefont{Y.~K.} \bibnamefont{Luo}},
  \bibinfo{author}{\bibfnamefont{C.~H.} \bibnamefont{Lee}},
  \bibinfo{author}{\bibfnamefont{M.~R.} \bibnamefont{Brenner}},
  \bibinfo{author}{\bibfnamefont{S.}~\bibnamefont{Rajan}},
  \bibinfo{author}{\bibfnamefont{J.~A.} \bibnamefont{Gupta}},
  \bibinfo{author}{\bibfnamefont{D.~W.} \bibnamefont{McComb}},
  \bibnamefont{et~al.}, \bibinfo{journal}{Nano Letters}
  \textbf{\bibinfo{volume}{18}}, \bibinfo{pages}{3125} (\bibinfo{year}{2018}),
  \urlprefix\url{https://doi.org/10.1021/acs.nanolett.8b00683}.

\bibitem[{\citenamefont{Vishkayi et~al.}(2020)\citenamefont{Vishkayi,
  Torbatian, Qaiumzadeh, and Asgari}}]{PhysRevMaterials.4.094004}
\bibinfo{author}{\bibfnamefont{S.~I.} \bibnamefont{Vishkayi}},
  \bibinfo{author}{\bibfnamefont{Z.}~\bibnamefont{Torbatian}},
  \bibinfo{author}{\bibfnamefont{A.}~\bibnamefont{Qaiumzadeh}},
  \bibnamefont{and} \bibinfo{author}{\bibfnamefont{R.}~\bibnamefont{Asgari}},
  \bibinfo{journal}{Phys. Rev. Materials} \textbf{\bibinfo{volume}{4}},
  \bibinfo{pages}{094004} (\bibinfo{year}{2020}),
  \urlprefix\url{https://link.aps.org/doi/10.1103/PhysRevMaterials.4.094004}.

\bibitem[{\citenamefont{Mukherjee et~al.}(2019)\citenamefont{Mukherjee,
  Chowdhury, Jana, and Voon}}]{Mukherjee_2019}
\bibinfo{author}{\bibfnamefont{T.}~\bibnamefont{Mukherjee}},
  \bibinfo{author}{\bibfnamefont{S.}~\bibnamefont{Chowdhury}},
  \bibinfo{author}{\bibfnamefont{D.}~\bibnamefont{Jana}}, \bibnamefont{and}
  \bibinfo{author}{\bibfnamefont{L.~C. L.~Y.} \bibnamefont{Voon}},
  \bibinfo{journal}{Journal of Physics: Condensed Matter}
  \textbf{\bibinfo{volume}{31}}, \bibinfo{pages}{335802}
  (\bibinfo{year}{2019}),
  \urlprefix\url{https://doi.org/10.1088/1361-648x/ab1fcf}.

\bibitem[{\citenamefont{Wang et~al.}(2020)\citenamefont{Wang, Wang, Liang, Ma,
  Xu, Liu, Zhang, Admasu, Cheong, Wang et~al.}}]{strain1}
\bibinfo{author}{\bibfnamefont{Y.}~\bibnamefont{Wang}},
  \bibinfo{author}{\bibfnamefont{C.}~\bibnamefont{Wang}},
  \bibinfo{author}{\bibfnamefont{S.-J.} \bibnamefont{Liang}},
  \bibinfo{author}{\bibfnamefont{Z.}~\bibnamefont{Ma}},
  \bibinfo{author}{\bibfnamefont{K.}~\bibnamefont{Xu}},
  \bibinfo{author}{\bibfnamefont{X.}~\bibnamefont{Liu}},
  \bibinfo{author}{\bibfnamefont{L.}~\bibnamefont{Zhang}},
  \bibinfo{author}{\bibfnamefont{A.~S.} \bibnamefont{Admasu}},
  \bibinfo{author}{\bibfnamefont{S.-W.} \bibnamefont{Cheong}},
  \bibinfo{author}{\bibfnamefont{L.}~\bibnamefont{Wang}}, \bibnamefont{et~al.},
  \bibinfo{journal}{Advanced Materials} \textbf{\bibinfo{volume}{32}},
  \bibinfo{pages}{2004533} (\bibinfo{year}{2020}),
  \urlprefix\url{https://onlinelibrary.wiley.com/doi/abs/10.1002/adma.202004533}.

\bibitem[{\citenamefont{Rold{\'{a}}n et~al.}(2015)\citenamefont{Rold{\'{a}}n,
  Castellanos-Gomez, Cappelluti, and Guinea}}]{Rold_n_2015}
\bibinfo{author}{\bibfnamefont{R.}~\bibnamefont{Rold{\'{a}}n}},
  \bibinfo{author}{\bibfnamefont{A.}~\bibnamefont{Castellanos-Gomez}},
  \bibinfo{author}{\bibfnamefont{E.}~\bibnamefont{Cappelluti}},
  \bibnamefont{and} \bibinfo{author}{\bibfnamefont{F.}~\bibnamefont{Guinea}},
  \bibinfo{journal}{Journal of Physics: Condensed Matter}
  \textbf{\bibinfo{volume}{27}}, \bibinfo{pages}{313201}
  (\bibinfo{year}{2015}),
  \urlprefix\url{https://doi.org/10.1088/0953-8984/27/31/313201}.

\bibitem[{\citenamefont{Chumak et~al.}(2015)\citenamefont{Chumak, Vasyuchka,
  Serga, and Hillebrands}}]{chumak2015magnon}
\bibinfo{author}{\bibfnamefont{A.~V.} \bibnamefont{Chumak}},
  \bibinfo{author}{\bibfnamefont{V.~I.} \bibnamefont{Vasyuchka}},
  \bibinfo{author}{\bibfnamefont{A.~A.} \bibnamefont{Serga}}, \bibnamefont{and}
  \bibinfo{author}{\bibfnamefont{B.}~\bibnamefont{Hillebrands}},
  \bibinfo{journal}{Nature Physics} \textbf{\bibinfo{volume}{11}},
  \bibinfo{pages}{453} (\bibinfo{year}{2015}).

\bibitem[{\citenamefont{Xing et~al.}(2019)\citenamefont{Xing, Qiu, Wang, Yao,
  Ma, Cai, Jia, Xie, and Han}}]{PhysRevX.9.011026}
\bibinfo{author}{\bibfnamefont{W.}~\bibnamefont{Xing}},
  \bibinfo{author}{\bibfnamefont{L.}~\bibnamefont{Qiu}},
  \bibinfo{author}{\bibfnamefont{X.}~\bibnamefont{Wang}},
  \bibinfo{author}{\bibfnamefont{Y.}~\bibnamefont{Yao}},
  \bibinfo{author}{\bibfnamefont{Y.}~\bibnamefont{Ma}},
  \bibinfo{author}{\bibfnamefont{R.}~\bibnamefont{Cai}},
  \bibinfo{author}{\bibfnamefont{S.}~\bibnamefont{Jia}},
  \bibinfo{author}{\bibfnamefont{X.~C.} \bibnamefont{Xie}}, \bibnamefont{and}
  \bibinfo{author}{\bibfnamefont{W.}~\bibnamefont{Han}},
  \bibinfo{journal}{Phys. Rev. X} \textbf{\bibinfo{volume}{9}},
  \bibinfo{pages}{011026} (\bibinfo{year}{2019}),
  \urlprefix\url{https://link.aps.org/doi/10.1103/PhysRevX.9.011026}.

\end{thebibliography}

\end{document}